\documentclass[a4paper,preprint,10pt]{elsarticle}
\usepackage{microtype}
\usepackage{amsfonts}
\usepackage{amsmath}
\usepackage[displaymath]{preview}

\usepackage{pgfplots}

\usepackage{bm}

\usepackage{subfig}

\usepackage{fixltx2e}
\usepackage{hyperref}

\usepackage{bbm}

\title{An optimal linear solver for the Jacobian~system of the extreme~type-II~Ginzburg--Landau~problem}

\author[ua]{N.~Schl\"omer}
\ead{nico.schloemer@ua.ac.be}

\author[ua]{W.~Vanroose}
\ead{wim.vanroose@ua.ac.be}

\address[ua]{Universiteit~Antwerpen, Departement~Wiskunde-Informatica, Middelheimlaan~1, 2020~Antwerp, Belgium}

\newcommand\revcolor{black}
\newcommand\rev[1]{#1}

\newcommand\A{\ensuremath{\bm{A}}}
\newcommand\x{\ensuremath{\bm{x}}}
\newcommand\ee{\ensuremath{\bm{e}}}
\newcommand\vv{\ensuremath{\bm{v}}}
\newcommand\w{\ensuremath{\bm{w}}}
\newcommand\0{\ensuremath{\bm{0}}}
\newcommand\B{\ensuremath{\bm{B}}}
\renewcommand\H{\ensuremath{\bm{H}}}
\newcommand\n{\ensuremath{\mathbf{n}}}
\newcommand\R{\ensuremath{\mathbb{R}}}

\newcommand\C{\ensuremath{\mathbb{C}}}
\newcommand\bn{\ensuremath{\bm{\nabla}}}
\newcommand\Dpsi{\ensuremath{\delta\psi}}

\renewcommand\i{\ensuremath{\textup{i}}}

\newcommand\GL{\ensuremath{\mathcal{G\!L}}}
\newcommand\dfn{\ensuremath{\mathrel{\mathop:}=}}
\newcommand\nfd{\ensuremath{=\mathrel{\mathop:}}}

\newcommand\igralnl[4]{\ensuremath{\int\nolimits_{#1}^{#2} #3 \, \mathrm{d} #4}}

\newcommand\tp{\ensuremath{\mathrm{T}}}

\newcommand\J{\ensuremath{J}}
\newcommand\K{\ensuremath{K}}
\newcommand\conj[1]{\ensuremath{\overline{#1}}}

\newcommand\relphantom[1]{\mathrel{\phantom{#1}}}

\newtheorem{thm}{Theorem}
\newtheorem{lem}[thm]{Lemma}
\newtheorem{cor}[thm]{Corollary}
\newdefinition{rmk}{Remark}
\newproof{pf}{Proof}

\DeclareMathOperator{\spn}{span}

\DeclareMathOperator{\amg}{AMG}

\newlength\figurewidth
\newlength\figureheight
\begin{document}
\begin{abstract}
This paper considers the extreme type-II
Ginzburg--Landau~equations, a nonlinear PDE model for describing the states of
a wide range of superconductors.
Based on properties of the Jacobian operator and an
AMG strategy, a preconditioned Newton--Krylov method is constructed.
After a finite-volume-type discretization, numerical experiments
are done for representative two- and three-dimensional domains.
Strong numerical evidence is provided that the
number of Krylov iterations is independent of the dimension $n$ of the
solution space, yielding an overall solver complexity of $O(n)$.
\end{abstract}

\begin{keyword}
Ginzburg--Landau equations \sep preconditioning \sep algebraic multigrid
\end{keyword}

\maketitle
\section{Introduction}
The nonlinear Schr\"odinger equation is used in many areas of science
and technology and describes, for example, the propagation of
solutions in fiber optics~\cite{taylor1992optical} and Bose--Einstein condensates in ultra-cold
traps~\cite{Bao2003318}. Prototypical for this type of models
is the Ginzburg--Landau problem, widely used to study
the state of both low- and
high-temperature superconductors. Due to its highly nonlinear nature, the involved energy landscape, and
the strong dependence of solutions on external conditions, numerical
simulations of the Ginzburg--Landau model have become an essential tool for
providing better insight into properties of
superconductivity phenomena.

The Ginzburg--Landau model has attracted wide
interest since its inception in the 1950s. In particular, the work on
the linearization by Abrikosov of the model around the upper critical
field is widely known \cite{abrikosov1957}.  The mathematical foundations
for the equilibrium Ginzburg--Landau models are well developed
\cite{sandier2007vortices,LD:1997:GLV} and a framework for finite
element and finite volume discretizations was provided \cite{DGP:1993:MAP}.
Different types of discretizations and numerical approximations of the
Ginzburg--Landau models have been developed since, all of which
subject to numerical simulations.

Throughout the physics literature, several methods for
solving the Ginzburg--Landau equations are described.
Used most prominently is a Gauss--Seidel-type iterative scheme
\cite{PhysRevLettSchweigert,PhysRevBMilosevic} that is readily
implemented, yet fails to converge for systems with physically
unstable vortex configurations.
Furthermore, it only yields linear convergence close to a solution.
In computational physics in general, the use of Newton--Krylov methods and
nonlinear multigrid schemes such as FAS is
widespread~\cite{knoll2004jacobian}.
Also, preconditioned Newton--Krylov methods are already applied
in other phase field models such as the Cahn--Hilliard equation \cite{yang2011}.
Initial efforts to apply
Newton--Krylov to the Ginzburg--Landau problem were taken in \cite{carey2010};
preconditioning is not discussed though.

\begin{figure}
\centering
\includegraphics{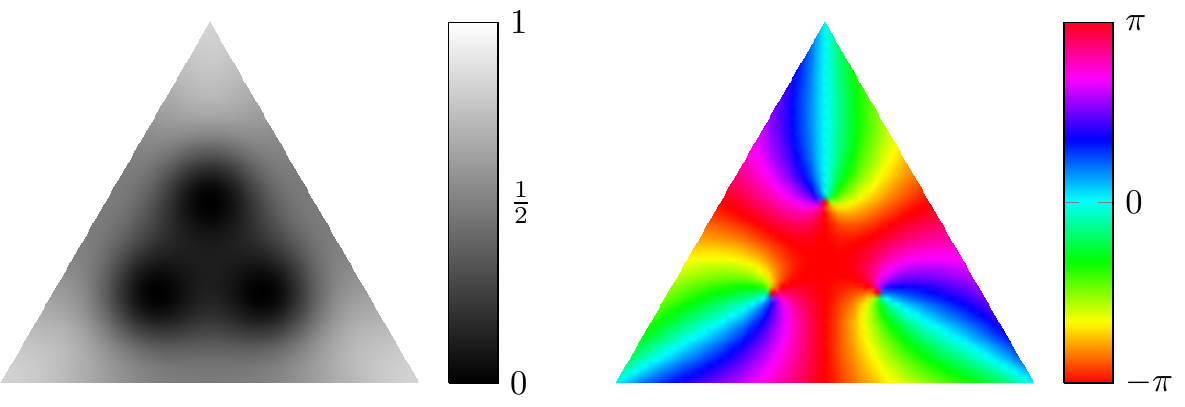}
\caption{Typical solution $\psi\colon\Omega\mapsto\C$ (displayed as $|\psi|^2$,
$\arg\psi$) of the extreme-type-II Ginzburg--Landau equations, (\ref{eq:GL}),
here for a flat triangular domain with circumradius 5 and the magnetic vector
potential $\A(x,y)=(-y/2,x/2)^\tp$. Three of the characteristic vortices
appear.}
\label{fig:example}
\end{figure}

An important research topic in the context of the Ginzburg--Landau equations
is the formation of vortex patterns in the solutions (see figures~\ref{fig:example},
\ref{fig:step22}, \ref{fig:step49}).
To understand the formation and dynamics of those patterns,
the tools of nonlinear systems analysis can be employed.
For example, numerical continuation
techniques help computing a family of solutions as a function of a problem parameter,
e.g., the strength of the externally applied magnetic field or
the electric current at one of the boundaries.
The main application of numerical parameter
continuation is the construction of a bifurcation diagram that
identifies the stability regions and the transition between stable and
unstable patterns marked by bifurcation points \cite{krauskopf2007}.
A systematic
bifurcation analysis of the patterns that appear in mesoscopic
superconductors is carried out for square-shaped domains in \cite{SAV:2012:NBS}.
The main
computational load in numerical continuation are the linear solves with
the Jacobian operator.
By the sheer number of unknowns,
this is particularly expensive for discretizations of
three-dimensional domains. It is thus required to develop
linear solvers for which the memory requirements
and the computational cost grows slowly with the number of unknowns.
To the knowledge of the authors, no linear scalable method for the
Ginzburg--Landau problem has been developed.
It is the goal of this paper to display that an AMG-preconditioned
Newton--Krylov method is a viable approach for the extreme-type-II
Ginzburg--Landau equations.

The remainder of the paper is organized as follows.
Section~\ref{sec:review} reviews the Ginzburg--Landau equations for
extreme-type-II superconductors;
section~\ref{sec:numericalmethods} is concerned with its linearization,
the Jacobian,  and discusses properties with respect to numerical algorithms.
While section~\ref{sec:disc} introduces the applied discretization
and shows that many important properties carry over from the continuous framework,
section~\ref{sec:algo} is concerned with the solution of the Jacobian
system and introduces a multigrid strategy.
The convergence behavior is explored through numerical experiments on
representative two- and three-dimensional domains.
The document concludes with a discussion of the obtained results.

\section{The Gibbs energy and the continuous Ginzburg--Landau problem}\label{sec:review}

For an open, bounded domain $\Omega\subset\R^3$
with a piecewise smooth boundary $\partial\Omega$, the Ginzburg--Landau
problem is usually stated as a minimization problem of the Gibbs
energy functional
\begin{equation}\label{eq:Gibbs}
\begin{split}
G(\psi,\A) - G_{\mathrm{n}}
&=
\xi\frac{|\alpha|^2}{\beta}
\int_{\Omega}
\Bigg[
-|\psi|^2
+ \frac{1}{2}|\psi|^4
+ \left\|-\i\bn\psi - \A\psi \right\|^2\\
&\relphantom{=}\phantom{\xi\frac{|\alpha|^2}{\beta}\int_{\Omega}\Bigg[}
+ \kappa^2 (\bn\times \A)^2
- 2\kappa^2 (\bn\times \A)\cdot \H_0
\Bigg]\, \mathrm{d}\Omega
\end{split}
\end{equation}
over $\psi\in H^2_{\C}(\Omega)$ and $\A\in H_{\R^n}^2(\Omega)$ \cite{DGP:1993:MAP}.
The scalar-valued function $\psi$ is commonly
referred to as \emph{order parameter}, $\A$ is
the magnetic vector potential corresponding to the total magnetic field.
The physical observables associated with the state $(\psi,\A)$ are the density
$\rho_{\text{C}}=|\psi|^2$ of the superconducting charge carriers (\emph{Cooper
pairs}) and the magnetic
field $\B=\bn\times\A$. The constant $G_{\mathrm{n}}$ represents the energy
associated with the entirely normal (non-superconducting) state.

The energy~(\ref{eq:Gibbs}) is presented in its dimensionless form,
and it depends upon the impinging magnetic field $\H_0$ and the
material parameters $\alpha, \beta, \lambda, \xi\in\R$.  The ratio
$\kappa\dfn\lambda/\xi$ of the penetration depth $\lambda$ (the length
scale at which the magnetic field penetrates the sample) and the
coherence length $\xi$ (the characteristic spatial scale of $\psi$)
determines the type of the superconductor: It is said to be of
\emph{type I} if $\kappa < 1/\sqrt{2}$, and of \emph{type II}
otherwise.  The two types behave fundamentally differently when
exposed to a magnetic field: \emph{Type~I} superconductors exhibit
alternating superconducting and nonsuperconducting regions,
while \emph{type~II}
superconductors show vortex patterns~\cite{huebener2001magnetic} (see figure~\ref{fig:example}).

Starting from the Gibbs energy and using standard calculus of variations, it is
possible to derive the Ginzburg--Landau equations \cite{DGP:1992:AAG}, a
boundary-value problem in the unknowns $\psi$ and $\A$.
As anticipated in the introduction, we will simplify the problem and
consider only the limit $\kappa\to\infty$ (\emph{extreme type-II
  superconductors}): this approximation gives satisfactory results for
all high-temperature superconductors which have large values of
$\kappa$ (typically $50<\kappa$).  In this case, the
Ginzburg--Landau equations decouple for $\psi$ and $\A$, such that
the magnetic vector potential $\A$ is given up to gauging
by the applied magnetic field $\H_0$ through
\[
\bn\times\A=\H_0 \quad\text{in } \R^3,
\]
and $\psi\in X$, $X\subseteq H_{\C}^2(\Omega)$, by
\begin{equation}\label{eq:GL}
0 = \GL(\psi)\dfn
\begin{cases}
\left(-\i\bn - \A\right)^2 \psi - \psi \left(1 - |\psi|^2\right) \quad \text{in } \Omega,  \\[3mm]
\n \cdot ( -\i\bn - \A) \psi \quad \text{on } \partial\Omega,
\end{cases}
\end{equation}
with \rev{$\n$ being the outer normal on $\partial\Omega$ and
$X\dfn\{\psi\in H_{\C}^2(\Omega): \GL(\psi,\A)\text{ bounded}\}$, i.e.,}
the natural energy space of (\ref{eq:Gibbs}).

As the domain is given in units of $\xi=\lambda/\kappa$,
the large-$\kappa$ limit implies $\lambda\gg\xi$ which
means that $\H_0$ is not disturbed by the magnetic field
induced by the electric charge density $\rho_C$.

Note that, for any given $\chi\in\R$,
\[
\GL(\exp(\i\chi)\psi)=\exp(\i\chi)\GL(\psi).
\]
Consequently, any given solution $\psi$ of the Ginzburg--Landau problem is really
just a representative of a whole set of solutions $[\psi]\dfn\{\exp(\i\chi)\psi\colon\chi\in\R\}$.
This expresses the fact that for superconducting states, the actual value of the
argument of $\psi$ is of no physical relevance: $|\psi|^2$ represents the observable.
As the complex argument of any coefficient does not play any role in the scalar multiplication,
$c [\psi] = |c| [\psi]$, $c\in\C$, $\psi\in X$,
it is natural to restrict the scalar field to $\R$. The inner product in the
vector space $X$ over the field $\R$ is
\begin{equation}\label{eq:scalar product}
\left\langle \phi, \psi\right\rangle_{\R}
\dfn \Re \langle \phi, \psi\rangle
= \Re\left( \int_\Omega \conj{\phi} \psi \right).
\end{equation}

\subsection{The Jacobian, the kinetic energy operator, and their properties}\label{sec:numericalmethods}

Equation (\ref{eq:GL}) is a nonlinear equation in $\psi$ and
hence classically suited for treatment with Newton's method. While there
were efforts to solve (\ref{eq:GL}) with a modified
algorithm \cite{KK:1995:VCT}, the generic approach of the full Newton
system is applied here for its attractive second-order convergence.
In this section, properties of the (continuous) Jacobian system
\begin{equation}\label{eq:Newton system}
\J(\psi) \Dpsi = -\GL(\psi)
\end{equation}
with
\begin{equation}\label{eq:jacobian}
\J(\psi)\varphi
\dfn \left( (-\i \bn - \A)^2  - 1 + 2|\psi|^2 \right) \varphi + \psi^2 \conj{\varphi}.
\end{equation}
will be discussed.
Note that $\J(\psi)$ is \rev{only linear if $X$ is defined as vector space
over the field $\R$}.

\paragraph{The kinetic energy operator}

Before analyzing the Jacobian operator $\J(\psi)$ as a whole, we will take a close
look at the part that is commonly referred to as the \emph{kinetic energy operator},
\begin{equation}\label{eq:kinetic energy operator}
\K\varphi \dfn (-\i\bn-\A)^2 \varphi.
\end{equation}
This operator is linear in $X$ and self-adjoint with respect to the ordinary $L^2(\Omega)$-inner product
(see \cite{SAV:2012:NBS}).
Consequently, all eigenvalues of $\K$ are real-valued.
Even more can be stated about its spectrum: From
\[
\int_{\Omega} \overline{\psi} (-\i\bn-\A)^2\varphi
= \int_{\Omega} \overline{(-\i\bn-\A)\psi} (-\i\bn-\A)\varphi
- \i \int_{\partial\Omega} \conj{\psi} \n\cdot (-\i\bn-\A)\varphi
\]
for all $\psi, \varphi\in L^2_{\C}(\Omega)$,
it follows that the kinetic energy operator
is positive-semidefinite over the subspace $\widetilde{X}\subseteq X$,
\[
\widetilde{X}\dfn\{ \psi\in X: \n\cdot (-\i\bn-\A)\rev{\psi} = 0 \text{ a.e.\ \rev{on} }\rev{\partial\Omega}\}.
\]
This is because for all $\psi\in \widetilde{X}$,
\[
\left\langle \psi, \K\psi \right\rangle_{L^2(\Omega)}
=
\int_{\Omega} \overline{\psi} \K \psi\,\mathrm{d}\Omega
= \int_{\Omega} \left\|(-\i\bn-\A)\psi\right\|^2 \,\mathrm{d}\Omega \ge 0.
\]
Moreover, the value of $0$ is attained if and only if
\[
(-\i\bn-\A)\psi = \rev{\0} \quad\text{a.e.\ on }\Omega,
\]
from which in turn follows that
\begin{multline}\label{eq:keo degenerate}
\rev{\0}
= \bn\times (-\i\bn\psi) - \bn\times(\A\psi)
= -\i (\bn\times \bn\psi) - (\bn\times\A)\psi - (\bn\psi)\times\A\\
= - \B\psi - \i (\A\psi)\times\A
= - \B \psi \quad\text{a.e.\ on } \Omega.
\end{multline}
Hence, only for vanishing magnetic fields $\B$, the kinetic energy operator $\K$
is actually degenerate.

An approximation for the smallest magnitude eigenvalue around the \rev{constant} zero-field $\A_0\equiv\0$
can be obtained by \rev{eigenvalue} perturbation.
Note that $\K(\A_0)$ is the Laplace operator with homogeneous Neumann boundary conditions,
so the smallest magnitude eigenvalue of $\K(\A_0)$ is $\lambda_0=0$, the
corresponding \rev{constant eigenfunction} $v_0\equiv 1$.
\rev{For the perturbed problem $(\K(\A_0)+\delta\K)(v_0+\delta v) = (\lambda_0+\delta\lambda)(v_0+\delta v)$,
one gets
\[
\delta\lambda\langle v_0, v_0\rangle = \langle v_0,(\delta K)v_0\rangle
              + \langle v_0,(\delta K-I\delta\lambda) \delta v\rangle,
\]
such that, in first-order approximation,
\[
\delta\lambda
\approx \frac{\langle v_0,(\delta K)v_0\rangle}{\langle v_0, v_0\rangle}
= \frac{\langle v_0,(\K(\A) - \K(\A_0))v_0\rangle}{\langle v_0, v_0\rangle}
\]
}
Noting that $\K(\A)v_0 = (-\i\bn-\A)^2 v_0 = \A^2 v_0$, this yields
\begin{equation}\label{eq:lambda0 cont}
\lambda
\approx \lambda_0 + \frac{\left\langle v_0, \A^2v_0 \right\rangle}{\langle v_0, v_0\rangle}
= |\Omega|^{-1} \int_{\Omega} \A^2.
\end{equation}

This shows  a lot more of the structure of the Jacobian operator $\J(\psi)$ already:
For any given $\psi\in X$, $\J(\psi)$ is the composition of a self-adjoint,
positive-(semi)definite operator and some reaction terms.

It is possible
to infer certain properties of $\J$ starting from here. From a numerical point of view,
insight into the adjointness and the spectrum of the operator will be highly
desirable.
The peculiar structure of $\J(\psi)$, acting on $\varphi$ and its pointwise complex conjugate,
together with the inner product~\eqref{eq:scalar product} in $X$, yield
\begin{lem}\label{lem:j self-adjoint}
  For any given $\psi\in L^2(\Omega)$, the Jacobian operator
  $\J(\psi)$ (\ref{eq:jacobian}) is linear and self-adjoint with
  respect to the inner product~\eqref{eq:scalar product}.
\end{lem}
\begin{pf}
See \cite{SAV:2012:NBS}.
\end{pf}

Now that the spectrum of $\J(\psi)$ is known to be a subset of $\R$ as well,
the natural question to ask is whether or not $\J(\psi)$ is generally
definite. Unfortunately, no such thing is true.
Quite the contrary: Note that, for any solution $\psi_{\mathrm{s}}$ of
(\ref{eq:GL}), we have
\begin{multline}\label{eq:jacobian nullspace}
\J(\psi_{\mathrm{s}}) (\i\psi_{\mathrm{s}})
=  \left[ (-\i \bn - \A)^2  - 1 + 2|\psi_{\mathrm{s}}|^2 \right] (\i\psi_{\mathrm{s}}) - \i \psi_{\mathrm{s}}^2  \conj{\psi_{\mathrm{s}}}\\
=  \left(1-|\psi_{\mathrm{s}}|^2\right)(\i\psi_{\mathrm{s}})  - \i\psi_{\mathrm{s}} + 2\i\conj{\psi_{\mathrm{s}}}\psi_{\mathrm{s}}^2  - \i \psi_{\mathrm{s}}^2  \conj{\psi_{\mathrm{s}}}
= 0,
\end{multline}
and hence $\spn\{\i\psi_{\mathrm{s}}\}\subseteq \ker\J(\psi_{\mathrm{s}})$.
This is a direct consequence of the fact that $\GL(\psi)$ (\ref{eq:GL}) is invariant
under the transformation $\widetilde{\psi}= \exp(\i\chi)\psi$
for any $\chi\in\R$.

Besides the fact that there is always a degenerate eigenvalue and that all
eigenvalues are real, not much more can be said about the spectrum; in general,
$\mathcal{J}(\psi)$ is indefinite. The definiteness depends entirely on the
state $\psi$; if $\psi$ is a solution to (\ref{eq:GL}), it
is said to be physically stable or unstable depending on whether or not $\mathcal{J}(\psi)$
has positive eigenvalues. Typically, solutions with relatively low energy
tend to be stable whereas solutions with relatively high energy tend to be unstable.
\section{Discretization in finite volumes and link variables}\label{sec:disc}

In recent years, the research in applications for superconductors has
taken strong interest in studying the effect of the
sample geometry on superconductivity phenomena, for example, of dents or holes
in a domain. Such geometries cannot be captured well by classical Cartesian staggered
grids \cite{SAV:2012:NBS}, so finite element and finite volume approaches have been
developed that incorporate properties of the continuous Ginzburg--Landau
equations such as the gauge invariance. In \cite{du2005numerical},
the method has been described for two-dimensional domains
and shall be described here in general terms.

\begin{figure}
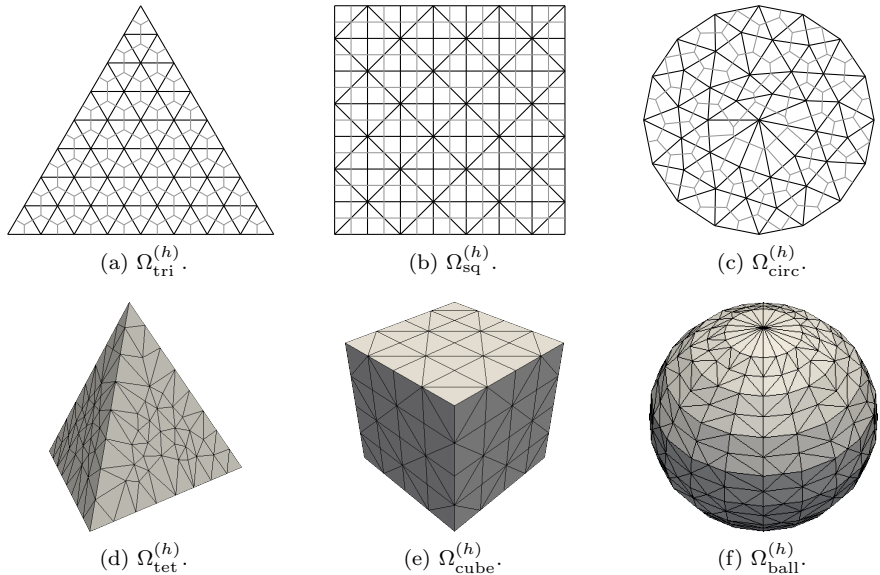

\setlength\figureheight{0.25\textwidth}

\caption{Triangulations of test domains, shown in rather coarse discretizations.
For the two-dimensional domains, the Voronoi regions $V_i$ as appearing
in~(\ref{eq:mass lumping}) are highlighted in light gray.}
\label{fig:test domains}
\end{figure}

Let $\x_j\in\R^d$, $d\in\{2,3\}$, $j\in\{1,\dots,n\}$ be a given set of discretization points at which
states~$\psi\in X$ will be approximated by $\psi^{(h)}_j\approx\psi(\x_j)$,
$\psi^{(h)}\in\C^n$.
Each discretization point $\x_j$ be equipped with its corresponding \emph{Voronoi region},
\[
V_j\dfn\{\x\in\R^d\colon \|\x-\x_j\| < \|\x-\x_k\|\:\forall k\neq j\}.
\]
The set $\{V_j\}_{j=1}^n$ is referred to as \emph{Voronoi tessellation}
corresponding to the \emph{generator set} $\{\x_j\}_{j=1}^n$. The dual
to a Voronoi tessellation consists of simplices and is referred
to as \emph{Delaunay triangulation} $\{T_i\}_{i=1}^m$ (see, e.g., figure~\ref{fig:test domains}).

For the domain $\Omega^{(h)}\dfn\bigcup_{j=1}^n V_j = \bigcup_{i=1}^m T_i$,
the significant part of the Gibbs energy (\ref{eq:Gibbs}) can be written as
\[
F(\psi,\A)
=
\underbrace{\sum_{i=1}^m \int_{T_i} \left\|-\i\bn\psi - \A\psi \right\|^2}_{\nfd F_1(\psi,\A)}
+
\underbrace{\sum_{j=1}^n \int_{V_j} \left(-|\psi|^2 + \frac{1}{2}|\psi|^4\right)}_{\nfd F_2(\psi)}.
\]
The second term, $F_2$, is readily discretized by mass lumping,
\begin{equation}\label{eq:mass lumping}
F_2^{(h)}(\psi^{(h)})
\dfn
\sum_{j=1}^n |V_j| \left(-|\psi^{(h)}_j|^2 + \frac{1}{2}|\psi^{(h)}_j|^4\right).
\end{equation}

For the discretization of $F_1$, we will first refer to a
technique \rev{for triangular meshes} in \cite{du2005numerical}\rev{, extended to arbitrary dimension here.
\begin{lem}
Let $\ee_i$, $i\in\{1,\dots,n\}$, with $n\dfn d(d+1)/2$ be the edges of a nondegenerate $d$-dimensional simplex. Then the symmetric rank-1 matrices $\{\ee_i \ee_i^\tp\}_{i=1}^n$
form a basis of the vector space of symmetric $d\times d$-matrices.
\end{lem}
\begin{pf}
The number $n$ of edges in a $d$-dimensional simplex coincides with the dimensionality
of the vector space of symmetric $d\times d$-matrices. Hence, only linear independence has to be
shown. Assume then that
\begin{equation}\label{eq:lin indep}
0_{d,d} = \sum_{i=1}^n \beta_i (\ee_i \ee_i^\tp)
\end{equation}
with some $\beta\in\R^n$. Since the simplex is not degenerate, there is a regular matrix $U$
that maps the edges $\{\ee_i\}_{i=1}^{n}$ onto the edges $\{\tilde{\ee}_i\}_{i=1}^{n}$ of the unit simplex.
With this, (\ref{eq:lin indep}) is equivalent to
\begin{equation}\label{eq:lin indep2}
0_{d,d}
= U \left( \sum_{i=1}^n \beta_i (\ee_i \ee_i^\tp) \right) U^\tp
= \sum_{i=1}^n \beta_i (U\ee_i (U\ee_i)^\tp)
= \sum_{i=1}^n \beta_i (\tilde{\ee}_i \tilde{\ee}_i^\tp).
\end{equation}
For the edges parallel to one of the axes, we have
$\tilde{\ee}_i \tilde{\ee}_i^\tp = e_k e_k^\tp$
with $e_k$ being the unit vector in $k$-direction.
For the edges between the two axes $k_1$, $k_2$, we have
$\tilde{\ee}_i \tilde{\ee}_i^\tp = (e_{k_1}-e_{k_2}) (e_{k_1}-e_{k_2})^\tp$.
As the matrix that belongs to edge between $k_1$, $k_2$ is the only
matrix with a nonzero entry at $(k_1, k_2)$ (namely $-1$), its
coefficient in (\ref{eq:lin indep2}) must be $0$.
Similarly, the same holds for all other coefficients, such that
(\ref{eq:lin indep}) can only be fulfilled of $\beta_i=0$ for all $i\in\{1,\dots,n\}$.
Hence, the matrices $\{\ee_i \ee_i^\tp\}_{i=1}^n$ are linearly independent.
\end{pf}
Since the $\{\ee_i \ee_i^\tp\}_{i=1}^n$ form a basis, there exists in particular \rev{a unique set of} coefficients
$a\in\R^n$, such that
\[
I_{d,d} = \sum_{\text{edges }\ee_i} a_i (\ee_i \ee_i^\tp).
\]
From this, we immediately conclude
}
\begin{cor}\label{cor:simplex}
Given a \rev{nondegenerate}  simplex $S\in\R^d$ with edges $\ee_{i,j}\dfn \x_i-\x_j$, $i,j\in\{1,\dots,d+1\}$, $i\neq j$,
there are coefficients $\alpha_{i,j}$ such that
\[\rev{
\int_{S} \left\|\bm{u}\right\|^2_2
= |S| \left\langle\bm{u}, \bm{u}\right\rangle_2
= \sum_{\text{edges }\ee_{i,j}} \alpha_{i,j} \left\langle \bm{u}, \ee_{i,j} \right\rangle_2 \, \left\langle \ee_{i,j}, \bm{u}\right\rangle_2
= \sum_{\text{edges }\ee_{i,j}} \alpha_{i,j} |\left\langle \ee_{i,j}, \bm{u}\right\rangle_2|^2}
\]
for any  $\bm{u}\in\C^d$.
\end{cor}
\rev{Given a simplex, one way of determining the edge coefficients $\alpha_{i,j}$ is to solve}
the symmetric and positive-definite linear equation system $M\alpha = b$ with
\[
M_{i,j} \dfn \left\langle \ee_i, \ee_j \right\rangle_2^2,\quad b_i \dfn |S| \left\|\ee_i\right\|_{\rev{2}}^2,
\]
where the edges are indexed subsequently.
\begin{rmk}
For triangles, the edge coefficients $\alpha_{i,j}$ are explicitly given by
\[
\alpha_{i,j}
= \frac{1}{2} \cot\theta_{i,j}
= \frac{1}{2} \frac{t_{i,j}}{\sqrt{1-t_{i,j}^2}},
\]
where $\theta_{i,j}$ is the angle opposing the edge $\ee_{i,j}$ \cite{du2005numerical},
and $t_{i,j}\dfn \left\langle\frac{\ee_{i,k}}{\|\ee_{i,k}\|},\frac{\ee_{j,k}}{\|\ee_{j,k}\|}\right\rangle_2$
with $k\notin\{i, j\}$.
\end{rmk}

With \rev{corollary}~\ref{cor:simplex} (and the coefficients $\alpha_{i,j}$ from there),
$F_1$ can be approximated by
\[
F_1(\psi,\A)\approx\widehat{F}_1(\psi,\A) \dfn \sum_{i=1}^m \sum_{\text{edges\,\,\,}\ee_{j,k}\text{\,\,of\,\,}T_i} \alpha^{(i)}_{j,k} |\ee_{j,k}\cdot (-\i\bn - \A)\psi(\overline{\x}_{j,k})|^2
\]
with $\overline{\x}_{j,k}\dfn \frac{1}{2}(\x_j + \x_k)$, or, more compactly,
\begin{equation}\label{eq:fhat}
\widehat{F}_1(\psi,\A)
= \sum_{\text{edges }\ee_{j,k}} \alpha_{j,k} |\ee_{j,k}\cdot (-\i\bn - \A)\psi(\overline{\x}_{j,k})|^2
\end{equation}
with the edge coefficients
\[
\alpha_{j,k} \dfn \sum_{\substack{\text{simplices } T_i\\\text{adjacent to edge }\ee_{j,k}}}\alpha_{j,k}^{(i)}.
\]

One could now do a finite difference approximation in the differential terms
of (\ref{eq:fhat}) to receive
a Gibbs energy defined over the discretized function space $\C^n$.
Note, however, that such naive discretization schemes of the momentum operator
$-\i\bn - \A$ lead to systems that preserve gauge invariance -- inherent
to the Ginzburg--Landau equations -- only up to a certain order in the spatial
discretization. It is hence customary to rewrite the momentum operator in
terms of variables that ensure preservation of gauge invariance for any
pointwise discretization.
Following \cite{KK:1995:VCT}, for any given normalized spatial direction $\vv$, let
\begin{equation}\label{eq:link variable}
U_{\vv}(\x) \dfn \exp\left( -\i \igralnl{\hat{\x}}{\x}{\vv\cdot\A(\w)}{\w}  \right),
\end{equation}
with arbitrary, fixed $\hat{\x}\in\x+\spn\{\vv\}$ (e.g., $\hat{\x}=\x-(\x\cdot\vv) \vv$).
\rev{Since $U_{\vv}(\x)$ sits on the unit circle, one has $\overline{U_{\vv}(\x)}\,U_{\vv}(\x)=1$,
and} with this
\begin{equation}\label{eq:link equality}
\conj{U}_{\vv} \vv\cdot\bn(U_{\vv}\psi)
\equiv  \conj{U}_{\vv} \left( -\i \vv\cdot\A U_{\vv} \psi + U_{\vv} \vv\cdot\bn\psi \right)
\equiv \i \vv\cdot( -\i \bn - \A ) \psi.
\end{equation}
Thus, $\widehat{F}_1$ can be written as
\[
\widehat{F}_1(\psi,\A)
= \sum_{\text{edges }\ee_{j,k}} \alpha_{j,k} \left|\conj{U_{\ee_{j,k}}}(\overline{\x}_{j,k}) \ee_{j,k}\cdot \bn(U_{\ee_{j,k}}\psi)(\overline{\x}_{j,k})\right|^2.
\]
Finite difference approximation finally yields the discretization
\[
\begin{split}
F_1^{(h)}(\psi^{(h)},\A)
&\dfn \sum_{\text{edges }\ee_{j,k}} \alpha_{j,k} \left|\conj{U_{\ee_{j,k}}}(\overline{\x}_{j,k}) \left(U_{\ee_{j,k}}(\x_j)\psi^{(h)}_j - U_{\ee_{j,k}}(\x_k)\psi^{(h)}_k\right)\right|^2\\
&=\sum_{\text{edges }\ee_{j,k}} \alpha_{j,k} \left|U_{j,k}\psi^{(h)}_j - \psi^{(h)}_k\right|^2
\end{split}
\]
with
\[
U_{j,k} \dfn \exp\left(-\i\igralnl{\x_k}{\x_j}{\ee_{j,k}\cdot\A(\w)}{\w}\right),
\]
often called \emph{link-variable} \cite{KK:1995:VCT}.
If $\A$ is known only at certain points of along the edges,
$U_{j,k}$ could again be approximated by a quadrature formula.

Finally, together with \eqref{eq:mass lumping}, the discrete Ginzburg--Landau energy functional is
defined as
\begin{equation}\label{eq:discr energy}
F^{(h)}(\psi^{(h)},\A) \dfn F_1^{(h)}(\psi^{(h)},\A) +  F_2^{(h)}(\psi^{(h)}).
\end{equation}
The standard Euler--Lagrange formalism now yields a necessary condition
for extremal points of the energy functional,
\[
\begin{split}
0 &=
2\Re\Bigg(
\sum_{\text{edges }\ee_{j,k}} \alpha_{j,k}
\left[
\left(\psi^{(h)}_j - U_{j,k}\psi^{(h)}_k\right) \conj{\delta\psi}^{(h)}_j
+
\left(\psi^{(h)}_k - \conj{U_{j,k}}\psi^{(h)}_j\right) \conj{\delta\psi}^{(h)}_k
\right]\\
&\relphantom{0}\phantom{2\Re\Bigg(}
-
\sum_{j=1}^n |V_j| \psi^{(h)}_j \left( 1-|\psi^{(h)}_j|^2 \right) \overline{\delta\psi^{(h)}_j}
\Bigg)
\qquad \forall \delta\psi^{(h)}\in\C^n.
\end{split}
\]
This is equivalent to the discrete Ginzburg--Landau equations,
\begin{equation}\label{eq:discr ginla}
\forall i\in\{1,\dots,n\}:
\quad 0 = \left(K^{(h)}\psi^{(h)}\right)_i - \psi^{(h)}_i\left(1-|\psi^{(h)}_i|^2\right),
\end{equation}
where the discrete kinetic energy operator $K^{(h)}$ is defined by
\begin{multline}\label{eq:discr kin energy}
\forall\phi^{(h)},\psi^{(h)}\in\C^n: \quad \left\langle \phi^{(h)}, K^{(h)}\psi^{(h)}\right\rangle
=\\
\sum_{\text{edges }\ee_{j,k}} \alpha_{j,k}
\left[
\left(\psi^{(h)}_j - U_{j,k}\psi^{(h)}_k\right) \conj{\phi}^{(h)}_j
+
\left(\psi^{(h)}_k - \conj{U_{j,k}}\psi^{(h)}_j\right) \conj{\phi}^{(h)}_k
\right]
\end{multline}
with the discrete inner product
\begin{equation}\label{eq:discr inner}
\left\langle\psi^{(h)}, \phi^{(h)}\right\rangle \dfn \sum_{i=1}^n |V_i|\,\conj{\psi}^{(h)}_i \phi^{(h)}_i.
\end{equation}

\begin{rmk}\label{rmk:dk}
In matrix form, the operator $K^{(h)}$ is
represented as a product $K^{(h)}=D^{-1}\widehat{K}$ of the diagonal
matrix $D^{-1}$, $D_{i,i}=|V_i|$, and a Hermitian matrix $\widehat{K}$.
\end{rmk}

The discretization (\ref{eq:discr ginla}) has several advantages,
starting with the fact that the boundary conditions
of the Ginzburg--Landau equations~(\ref{eq:GL}) are naturally contained.
Also note
that the discrete kinetic energy operator~(\ref{eq:discr kin energy})
coincides, up to the terms $U_{j,k}$, with the discretization of the
Laplace operator with homogeneous Neumann boundary conditions.
Similarly, it has a number of desirable properties
that will make the iterative solution of the Jacobian system easier.

\begin{lem}
The discrete kinetic energy operator $K^{(h)}$ (\ref{eq:discr kin energy})
is self-adjoint with respect to the discrete inner product~(\ref{eq:discr inner}).
\end{lem}
\begin{pf}
Let $\phi^{(h)},\psi^{(h)} \in\C^n$. Then
\[
\begin{split}
\left\langle\phi^{(h)}, K^{(h)}\psi^{(h)} \right\rangle
&= \sum_{\text{edges }\ee_{j,k}} \alpha_{j,k} \left[ \left(\psi^{(h)}_j - U_{j,k}\psi^{(h)}_k\right) \conj{\phi}^{(h)}_j + \left(\psi^{(h)}_k - \conj{U_{j,k}}\psi^{(h)}_j\right) \conj{\phi}^{(h)}_k \right]\\
&= \sum_{\text{edges }\ee_{j,k}} \alpha_{j,k} \left[ \conj{\left(\phi^{(h)}_j - U_{j,k}\phi^{(h)}_k\right)} \psi^{(h)}_j + \conj{\left(\phi^{(h)}_k - \conj{U_{j,k}}\phi^{(h)}_j\right)} \psi^{(h)}_k \right]\\
&= \left\langle K^{(h)}\phi^{(h)}, \psi^{(h)} \right\rangle.
\end{split}
\]
\end{pf}

\begin{lem}\label{lem:discr K semidef}
The discrete kinetic energy operator $K^{(h)}$ (\ref{eq:discr kin energy}) is positive-semidefinite.
\end{lem}
\begin{pf}
Let $\psi^{(h)}\in\C^n$. Then
\[
\begin{split}
\left\langle \psi^{(h)}, K^{(h)} \psi^{(h)} \right\rangle
&= \sum_{\text{edges }\ee_{j,k}} \alpha_{j,k} \left[ \left(\psi^{(h)}_j - U_{j,k}\psi^{(h)}_k\right) \conj{\psi}^{(h)}_j + \left(\psi^{(h)}_k - \conj{U_{j,k}}\psi^{(h)}_j\right) \conj{\psi}^{(h)}_k \right]\\
&= \sum_{\text{edges }\ee_{j,k}} \alpha_{j,k} \left[ \psi^{(h)}_j \conj{\psi}^{(h)}_j - U_{j,k}\psi^{(h)}_k\conj{\psi}^{(h)}_j + \psi^{(h)}_k\conj{\psi}^{(h)}_k - \overline{U_{j,k}}\psi^{(h)}_j \conj{\psi}^{(h)}_k\right].
\end{split}
\]
Noting that $U_{j,k}\overline{U_{j,k}}=1$, this yields
\begin{equation}\label{eq:posdef}
\left\langle \psi^{(h)}, K^{(h)} \psi^{(h)} \right\rangle
= \sum_{\text{edges }\ee_{j,k}} \alpha_{j,k} \left|\psi^{(h)}_j - U_{j,k}\psi^{(h)}_k\right|^2 \ge 0.
\end{equation}
\end{pf}

For $\A_0\equiv\0$, the state $\widehat{\psi}^{(h)}\equiv 1$ is obviously an \rev{eigenvector} of $K^{(h)}$ with
the eigenvalue $0$.
Equation~(\ref{eq:posdef}) also delivers an approximation $\tilde{\lambda}_0$ for the smallest-magnitude
eigenvalue for perturbations of $\A_0$, namely
\[
\begin{split}
\tilde{\lambda}_0
= \frac{\left\langle \widehat{\psi}^{(h)}, K^{(h)} \widehat{\psi}^{(h)} \right\rangle}{\left\langle \widehat{\psi}^{(h)}, \widehat{\psi}^{(h)} \right\rangle}
&= |\Omega^{(h)}|^{-1} \sum_{\text{edges }\ee_{j,k}} \alpha_{j,k} \left|1 - U_{j,k}\right|^2\\
&= |\Omega^{(h)}|^{-1} \sum_{\text{edges }\ee_{j,k}} \alpha_{j,k} \left|2\sin \left(\frac{\arg U_{j,k}}{2}\right)\right|^2,
\end{split}
\]
or, in first approximation,
\begin{equation}\label{eq:smallest lambda}
\begin{split}
\tilde{\lambda}_0
&\approx \rev{|\Omega^{(h)}|^{-1} \sum_{\text{edges }\ee_{j,k}} \alpha_{j,k} \left|\arg U_{j,k}\right|^2}\\
&= |\Omega^{(h)}|^{-1} \sum_{\text{edges }\ee_{j,k}} \alpha_{j,k} \left|\igralnl{\x_k}{\x_j}{\ee_{j,k}\cdot\A(\w)}{\w}\right|^2.
\end{split}
\end{equation}
Compare this with the corresponding continuous expression~(\ref{eq:lambda0 cont}).

Completely analogous to the results for the continuous Jacobian operator $\J(\psi)$,
the discrete Jacobian operator
\[
\left(J^{(h)}(\psi^{(h)})\phi^{(h)}\right)_i \dfn
\left(K^{(h)}\phi^{(h)}\right)_i + \left(-1+2|\psi^{(h)}_i|^2\right)\phi^{(h)}_i + (\psi^{(h)}_i)^2\conj{\phi}^{(h)}_i
\]
of (\ref{eq:discr ginla}) is self-adjoint with respect to the inner product
\begin{equation}\label{eq:disc inner2}
\left\langle\phi^{(h)},\psi^{(h)}\right\rangle_{\R} \dfn \Re\left\langle\phi^{(h)},\psi^{(h)}\right\rangle.
\end{equation}
Like the continuous Jacobian operator $J(\psi)$, $J^{(h)}(\psi^{(h)})$ also has
a nontrivial kernel if $\psi^{(h)}$ is a solution to the problem.
While in the Newton process, the Jacobian system will never need to be solved in
exactly a solution, states close to a solution might introduce numerical difficulties
when nearly-singular systems need to be solved. Techniques for this situation
include adding phase conditions \cite{SAV:2012:NBS} or deflation methods.

\begin{rmk}\label{rmk:isometry}
Note that there is a vector space isometry of $\C^n$ as vector space over the field $\R$
and $\R^{2n}$ with its natural inner product: For all $\phi^{(h)}, \psi^{(h)}\in\C^n$,
\[
\left\langle
\begin{pmatrix}
\Re\phi^{(h)}\\
\Im\phi^{(h)}\\
\end{pmatrix},
\begin{pmatrix}
\Re\psi^{(h)}\\
\Im\psi^{(h)}\\
\end{pmatrix}
\right\rangle
=
\left\langle
\Re\phi^{(h)}, \Re\psi^{(h)}
\right\rangle
+
\left\langle
\Im\phi^{(h)}, \Im\psi^{(h)}
\right\rangle
=
\left\langle
\phi^{(h)}, \psi^{(h)}
\right\rangle_{\R}.
\]
This is relevant in practice if
the complex-valued original problem~(\ref{eq:discr ginla})
in $\C^n$ is implemented in terms of $\R^{2n}$. Using the natural inner
product in this space
will yield the expected results without having to take particular care of
$\left\langle\cdot,\cdot\right\rangle_{\R}$.
\end{rmk}
\section{Algorithm and numerical results}\label{sec:algo}

For real-world three-dimensional domains,
the solution of the discrete equivalent of the Jacobian system~(\ref{eq:Newton system})
will have too many unknowns for
black-box strategies such as $LU$-decompositions to fit into memory.
Exploiting the sparsity structure of the operator is crucial, and hence Krylov
subspace methods are an attractive alternative.
The choice of the appropriate Krylov subspace method depends on the spectral
properties of the operator.
Its self-adjointness (see lemma~\ref{lem:j self-adjoint} and its discrete equivalent)
make it possible for symmetric Krylov subspace methods
to efficiently solve the linear system
if adapted for the inner product~\eqref{eq:disc inner2} (see also remark~\ref{rmk:isometry}).
This avoids the larger memory-requirements that come with asymmetric solvers such as GMRES.
Furthermore, as $J^{(h)}(\psi^{(h)})$ is generally indefinite (depending on
$\psi^{(h)}$) and the number of negative eigenvalues can be large,
CG \rev{may be unsuitable} as a solver.
While it is known to perform well for indefinite problems when
the number of negative eigenvalues is not too large \cite{van2003iterative},
convergence can be irregular.
In contrast, MINRES is designed to deal with indefinite systems and is hence
\rev{a more suitable choice.}

\subsection{Preconditioning} \label{sec:preconditioning}
As the main computational effort of the nonlinear solver flows into the
linear solves of the Jacobian system, and the complexity of the linear solve
usually grows faster than linearly with the number of unknowns in the system,
it is crucial to explore the possibilities of accelerating the Krylov solver
using an appropriate preconditioner.
Given the results of
section~\ref{sec:numericalmethods}, we will evaluate the use of approximate
inverses of the operator
\[
P^{(h)}(\psi^{(h)}) \dfn K^{(h)} + 2|\psi^{(h)}|^2
\]
as a preconditioner for $J^{(h)}(\psi^{(h)})$.
The operator $P^{(h)}(\psi^{(h)})$ is obviously self-adjoint with respect to the standard
discrete inner product~(\ref{eq:discr inner}) and positive-semidefinite.
From lemma~\ref{lem:discr K semidef}, we can conclude that it is even
strictly positive-definite except for the uninteresting case $\psi^{(h)}\equiv 0$,
$\A\equiv \rev{\0}$.
Moreover, $K^{(h)}$ is derived from a geometric discretization, and its sparsity structure
coincides with that of the Laplacian with homogeneous Neumann boundary
conditions.
This makes the inversion of $P^{(h)}(\psi^{(h)})$ a suitable target for
algebraic-multigrid (AMG) strategies which are known to yield optimal convergence
behavior in the sense that the number of iterations required to reach a
certain stopping criterion is independent
of the number of unknowns in the system.
Furthermore, AMG methods are memory-efficient and scale well
in parallel computing environments \cite{Schroder:2012:SAS,BO:2008:AMK,ml-guide}.
The only nonstandard circumstance here is the fact that the matrix entries are complex-valued.
Difficulties in this area, however, were discussed and treated in
\cite{maclachlan:1548}.
\rev{\begin{rmk}
The operator $Q^{(h)}(\psi^{(h)})$ defined by
\[
\begin{split}
Q^{(h)}(\psi^{(h)}) \phi^{(h)}
&\dfn \left(J^{(h)}(\psi^{(h)}) + I\right) \phi^{(h)}\\
&= \left(K^{(h)} + 2|\psi^{(h)}|^2\right)\phi^{(h)} + (\psi^{(h)})^2 \overline{\phi^{(h)}} \quad \forall\phi^{(h)}\in\C^n
\end{split}
\]
is obviously self-adjoint and also positive-definite since
\[
\begin{split}
&\left\langle \phi^{(h)}, 2|\psi^{(h)}|^2 \phi^{(h)} + (\psi^{(h)})^2 \overline{\phi}^{(h)} \right\rangle_\R\\
&= \Re\left[\sum_{i=1}^n |V_i| \left(\overline{\phi}^{(h)}_i \cdot 2|\psi_i^{(h)}|^2 \phi^{(h)}_i + \overline{\phi}^{(h)}_i \cdot (\psi^{(h)}_i)^2 \overline{\phi}^{(h)}_i\right)\right]\\
&=  \sum_{i=1}^n |V_i| \left(2\left|\psi_i^{(h)}\overline{\phi}^{(h)}_i\right|^2 + \Re\left(\psi^{(h)}_i\overline{\phi}^{(h)}_i\right)^2 \right)\\
&> 0.
\end{split}
\]
It would hence also be a candidate for a good preconditioner. However, unlike $P^{(h)}$, it cannot be represented as a matrix and is thus not suitable for solution with AMG.
\end{rmk}}

Note that the
operator $\amg_k(A,b)$, defined by $k$ AMG cycles applied to a
Hermitian problem $Ax=b$, is again Hermitian. With remark~\ref{rmk:dk} (and $D$ from there),
we have that
\[
\left(P^{(h)}(\psi^{(h)})\right)^{-1} = \left(\widehat{K}+2D|\psi^{(h)}|^2\right)^{-1}D,
\]
so the approximate inverse of $P^{(h)}(\psi^{(h)})$,
\begin{equation}\label{eq:prec2}
R_k^{(h)}(\psi^{(h)}) = \amg_k\left(\widehat{K}+2D|\psi^{(h)}|^2,D\cdot\right),
\end{equation}
is self-adjoint with respect to the standard discrete inner product~(\ref{eq:discr inner}).

We will now explore this idea through numerical experiments, for the preconditioners
$R_1^{(h)}(\psi^{(h)})$ and
\begin{equation}\label{eq:prec1}
R^{(h)}_{\infty}(\psi^{(h)})
\dfn (P^{(h)}(\psi^{(h)}))^{-1}
= \left(\widehat{K}+2D|\psi^{(h)}|^2\right)^{-1} D,
\end{equation}
where $\widehat{K}+2D|\psi^{(h)}|^2$ is inverted numerically with high accuracy.

\rev{It is notoriously difficult to rigorously characterize the spectrum of
the Jacobian operator of the Ginzburg--Landau problem,
and the situation is similar for the preconditioned operator.}
Nevertheless,
if $(\lambda,\phi^{(h)})$ is an eigenvalue/eigenvector pair of the preconditioned
operator $R^{(h)}_{\infty}(\psi^{(h)})J^{(h)}(\psi^{(h)})$, i.e.,
\[
J^{(h)}(\psi^{(h)}) \phi^{(h)} = \lambda P^{(h)}(\psi^{(h)}) \phi^{(h)}.
\]
one gets
\begin{equation}\label{eq:lambda}
\lambda
= \frac{\left\langle \phi^{(h)}, J^{(h)}(\psi^{(h)}) \phi^{(h)}\right\rangle_{\R}}{\left\langle\phi^{(h)}, P^{(h)}(\psi^{(h)}) \phi^{(h)}\right\rangle_{\R}}
= 1 + \frac{\Re\langle(\phi^{(h)})^2,(\psi^{(h)})^2\rangle - 1}{\left\langle\phi^{(h)}, K^{(h)}\phi^{(h)}\right\rangle + 2\left\langle|\phi^{(h)}|^2,|\psi^{(h)}|^2\right\rangle}.
\end{equation}
In case $|\psi^{(h)}|\gg 1$ (which can happen during the Newton iteration),
the eigenvalues cluster around
$1 \pm \frac{1}{2}$ (depending on the sign of $\Re\langle(\phi^{(h)})^2,(\psi^{(h)})^2\rangle$), so the preconditioned
problem can be expected to be solved in a small number of Krylov iterations.
Noting that solutions $\psi$ of the Ginzburg--Landau equations (\ref{eq:GL}) fulfill $|\psi|<1$
pointwise, though, (\ref{eq:lambda}) unfortunately gives little insight in the behavior
close to a solution.
\rev{The same is true for the bounds gained from
estimating the denominator term $\left\langle\phi^{(h)}, K^{(h)}\phi^{(h)}\right\rangle$
with the help of the smallest eigenvalue approximation
for weak fields (\ref{eq:smallest lambda}).}

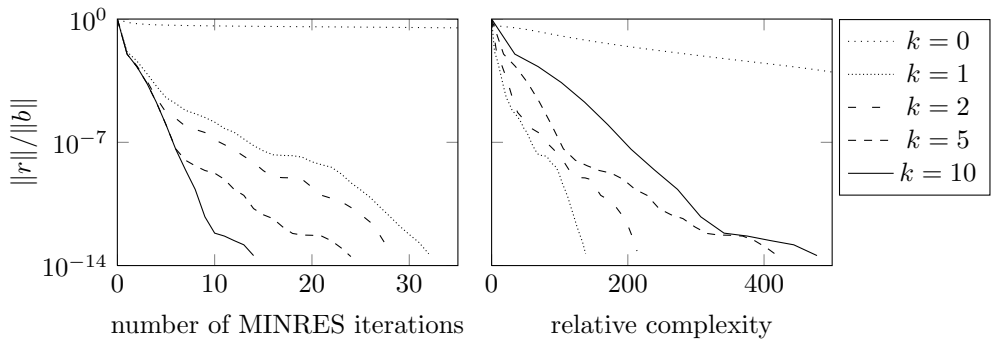
\begin{figure}
\setlength\figurewidth{0.5\textwidth}
\setlength\figureheight{0.8\figurewidth}
\centering
\subfloat{%
\begin{tikzpicture}[baseline]

\begin{semilogyaxis}[
xmin=0, xmax=35,
ymin=1e-14, ymax=1,
axis on top,
xlabel={number of MINRES iterations},
ylabel={$\|r\|/\|b\|$},
height=\figureheight,
width=\figurewidth
]
\addplot [\revcolor,dotted]
coordinates {
(0,1)  (1,0.644807)  (2,0.546145)  (3,0.49742)  (4,0.468268)  (5,0.453679)  (6,0.441702)  (7,0.429563)  (8,0.422672)  (9,0.414101)  (10,0.40827)  (11,0.402224)  (12,0.396963)  (13,0.392368)  (14,0.387608)  (15,0.383784)  (16,0.37946)  (17,0.376034)  (18,0.372062)  (19,0.368829)  (20,0.365125)  (21,0.36197)  (22,0.358459)  (23,0.355314)  (24,0.351929)  (25,0.348753)  (26,0.345443)  (27,0.342207)  (28,0.33893)  (29,0.335618)  (30,0.332342)  (31,0.328945)  (32,0.325644)  (33,0.322158)  (34,0.318815)  (35,0.31524)
};

\addplot [densely dotted]
coordinates {
(0,1) (1,0.0139658466039531) (2,0.0041668541398281) (3,0.000847698156070637) (4,0.000158136379991418) (5,3.47168021764028e-05) (6,1.58657865524238e-05) (7,6.49958511965696e-06) (8,4.22015623115358e-06) (9,2.57511691325517e-06) (10,1.40125091697247e-06) (11,6.16093768733579e-07) (12,3.71781289191082e-07) (13,1.53726633613046e-07) (14,6.52565438202897e-08) (15,3.28778752255404e-08) (16,1.95846870722275e-08) (17,1.79699504778845e-08) (18,1.7403260955763e-08) (19,1.43558186307187e-08) (20,7.78242754790345e-09) (21,5.32391815822228e-09) (22,3.83731158071241e-09) (23,1.62082917304287e-09) (24,5.07374790297855e-10) (25,1.98648892242197e-10) (26,7.32833323817348e-11) (27,2.08190482596783e-11) (28,6.22919571436964e-12) (29,1.76420320026259e-12) (30,5.82468868624697e-13) (31,2.04945468106169e-13) (32,4.66347260525612e-14)
};

\addplot [loosely dashed]
coordinates {
(0,1) (1,0.0105245937251093) (2,0.00219980728657953) (3,0.00033474466731038) (4,3.41878035950007e-05) (5,5.00048011464107e-06) (6,2.29402673953959e-06) (7,8.17760436119531e-07) (8,5.40246140746962e-07) (9,3.60361513215292e-07) (10,1.40351193592656e-07) (11,7.37327241828337e-08) (12,3.76840067690339e-08) (13,1.62467035686949e-08) (14,7.20099839149835e-09) (15,2.46373963882223e-09) (16,1.07636865373654e-09) (17,9.82222933846704e-10) (18,9.54419770641204e-10) (19,7.85662451050229e-10) (20,2.81235492222685e-10) (21,9.75578758138094e-11) (22,6.58084475902612e-11) (23,4.2905015390462e-11) (24,2.00802355450872e-11) (25,8.14921396691505e-12) (26,2.88791740160256e-12) (27,5.07567626264034e-13) (28,6.68390795775875e-14)
};

\addplot [dashed]
coordinates {
(0,1) (1,0.0102042983778133) (2,0.00199633712700245) (3,0.000247919901412568) (4,1.91691100717472e-05) (5,1.00340811333683e-06) (6,4.2766858142064e-08) (7,9.60396897586651e-09) (8,4.63042467611712e-09) (9,2.50863808804697e-09) (10,1.77144525432458e-09) (11,7.94439552478128e-10) (12,2.11831330952732e-10) (13,9.7102397306046e-11) (14,1.7146375391556e-11) (15,7.68907478564174e-12) (16,4.64025568958641e-12) (17,1.5596522324084e-12) (18,6.68147643877975e-13) (19,5.41368350608748e-13) (20,5.25496491148563e-13) (21,4.7110174511416e-13) (22,2.5693445967383e-13) (23,1.00997280049704e-13) (24,3.24515601161827e-14)
};

\addplot [black]
coordinates {
(0,1) (1,0.0102031443864273) (2,0.00199595424384562) (3,0.000247782434939009) (4,1.91485716863828e-05) (5,1.00149445993677e-06) (6,4.04356140970216e-08) (7,2.73590064131848e-09) (8,2.12281803016104e-10) (9,5.86538701950592e-12) (10,6.95539829793314e-13) (11,4.74299006171461e-13) (12,2.61968391539119e-13) (13,1.51058975156e-13) (14,3.69633947899524e-14)
};

\end{semilogyaxis}

\end{tikzpicture}}
\subfloat{%
\begin{tikzpicture}[baseline]

\begin{semilogyaxis}[
xmin=0, xmax=500,
ymin=1e-14, ymax=1,
axis on top,
xlabel={\rev{relative complexity}},
yticklabels={},
height=\figureheight,
width=\figurewidth,
legend style={
at={(1.02,1)},
anchor=north west
},
]
\addplot [\revcolor,dotted]
coordinates {
(0,1)  (1,0.644807)  (2,0.546145)  (3,0.49742)  (4,0.468268)  (5,0.453679)  (6,0.441702)  (7,0.429563)  (8,0.422672)  (9,0.414101)  (10,0.40827)  (11,0.402224)  (12,0.396963)  (13,0.392368)  (14,0.387608)  (15,0.383784)  (16,0.37946)  (17,0.376034)  (18,0.372062)  (19,0.368829)  (20,0.365125)  (21,0.36197)  (22,0.358459)  (23,0.355314)  (24,0.351929)  (25,0.348753)  (26,0.345443)  (27,0.342207)  (28,0.33893)  (29,0.335618)  (30,0.332342)  (31,0.328945)  (32,0.325644)  (33,0.322158)  (34,0.318815)  (35,0.31524)  (36,0.311845)  (37,0.308187)  (38,0.304733)  (39,0.300999)  (40,0.297486)  (41,0.293687)  (42,0.290118)  (43,0.286266)  (44,0.282649)  (45,0.278759)  (46,0.275102)  (47,0.271189)  (48,0.267503)  (49,0.263584)  (50,0.259883)  (51,0.255972)  (52,0.252268)  (53,0.248382)  (54,0.24469)  (55,0.240841)  (56,0.237174)  (57,0.233377)  (58,0.229747)  (59,0.226014)  (60,0.222433)  (61,0.218773)  (62,0.215251)  (63,0.211673)  (64,0.20822)  (65,0.204731)  (66,0.201354)  (67,0.19796)  (68,0.194666)  (69,0.191371)  (70,0.188164)  (71,0.184971)  (72,0.181855)  (73,0.178766)  (74,0.175744)  (75,0.172759)  (76,0.169832)  (77,0.166951)  (78,0.164121)  (79,0.161343)  (80,0.158608)  (81,0.155931)  (82,0.153292)  (83,0.150715)  (84,0.148169)  (85,0.145689)  (86,0.143235)  (87,0.14085)  (88,0.138486)  (89,0.136191)  (90,0.133915)  (91,0.131709)  (92,0.129518)  (93,0.127396)  (94,0.125287)  (95,0.123248)  (96,0.121218)  (97,0.119257)  (98,0.117304)  (99,0.115419)  (100,0.113539)  (101,0.111726)  (102,0.109917)  (103,0.108173)  (104,0.106432)  (105,0.104755)  (106,0.103079)  (107,0.101465)  (108,0.099852)  (109,0.0982982)  (110,0.0967452)  (111,0.0952496)  (112,0.093754)  (113,0.0923138)  (114,0.0908733)  (115,0.0894861)  (116,0.0880984)  (117,0.0867619)  (118,0.0854247)  (119,0.0841367)  (120,0.082848)  (121,0.0816064)  (122,0.0803641)  (123,0.0791671)  (124,0.0779691)  (125,0.0768146)  (126,0.0756593)  (127,0.0745456)  (128,0.0734311)  (129,0.0723565)  (130,0.0712812)  (131,0.0702441)  (132,0.0692063)  (133,0.0682051)  (134,0.0672034)  (135,0.0662366)  (136,0.0652696)  (137,0.0643359)  (138,0.063402)  (139,0.0625)  (140,0.0615981)  (141,0.0607266)  (142,0.0598553)  (143,0.0590131)  (144,0.0581712)  (145,0.0573572)  (146,0.0565436)  (147,0.0557566)  (148,0.0549703)  (149,0.0542093)  (150,0.0534491)  (151,0.0527131)  (152,0.0519781)  (153,0.0512661)  (154,0.0505554)  (155,0.0498666)  (156,0.0491791)  (157,0.0485126)  (158,0.0478475)  (159,0.0472024)  (160,0.0465589)  (161,0.0459346)  (162,0.0453118)  (163,0.0447073)  (164,0.0441046)  (165,0.0435193)  (166,0.0429358)  (167,0.042369)  (168,0.041804)  (169,0.041255)  (170,0.0407079)  (171,0.040176)  (172,0.0396462)  (173,0.0391307)  (174,0.0386175)  (175,0.038118)  (176,0.0376207)  (177,0.0371366)  (178,0.0366547)  (179,0.0361854)  (180,0.0357183)  (181,0.0352633)  (182,0.0348105)  (183,0.0343692)  (184,0.0339302)  (185,0.0335022)  (186,0.0330765)  (187,0.0326613)  (188,0.0322484)  (189,0.0318455)  (190,0.031445)  (191,0.031054)  (192,0.0306654)  (193,0.0302858)  (194,0.0299087)  (195,0.0295402)  (196,0.0291742)  (197,0.0288164)  (198,0.028461)  (199,0.0281136)  (200,0.0277685)  (201,0.027431)  (202,0.0270959)  (203,0.0267679)  (204,0.0264424)  (205,0.0261237)  (206,0.0258075)  (207,0.0254977)  (208,0.0251904)  (209,0.0248893)  (210,0.0245906)  (211,0.0242978)  (212,0.0240074)  (213,0.0237227)  (214,0.0234403)  (215,0.0231634)  (216,0.0228888)  (217,0.0226194)  (218,0.0223522)  (219,0.0220901)  (220,0.0218301)  (221,0.021575)  (222,0.0213221)  (223,0.0210737)  (224,0.0208275)  (225,0.0205857)  (226,0.0203461)  (227,0.0201106)  (228,0.0198772)  (229,0.0196479)  (230,0.0194206)  (231,0.0191972)  (232,0.0189758)  (233,0.0187581)  (234,0.0185425)  (235,0.0183303)  (236,0.0181202)  (237,0.0179134)  (238,0.0177086)  (239,0.017507)  (240,0.0173074)  (241,0.0171108)  (242,0.0169162)  (243,0.0167246)  (244,0.0165348)  (245,0.0163479)  (246,0.0161628)  (247,0.0159804)  (248,0.0157999)  (249,0.015622)  (250,0.0154459)  (251,0.0152723)  (252,0.0151005)  (253,0.0149311)  (254,0.0147635)  (255,0.0145981)  (256,0.0144345)  (257,0.0142731)  (258,0.0141134)  (259,0.0139559)  (260,0.0138)  (261,0.0136461)  (262,0.0134939)  (263,0.0133437)  (264,0.0131951)  (265,0.0130484)  (266,0.0129033)  (267,0.0127601)  (268,0.0126184)  (269,0.0124784)  (270,0.01234)  (271,0.0122034)  (272,0.0120682)  (273,0.0119347)  (274,0.0118026)  (275,0.0116722)  (276,0.0115432)  (277,0.0114158)  (278,0.0112898)  (279,0.0111653)  (280,0.0110422)  (281,0.0109205)  (282,0.0108003)  (283,0.0106814)  (284,0.0105639)  (285,0.0104478)  (286,0.0103329)  (287,0.0102194)  (288,0.0101073)  (289,0.00999635)  (290,0.00988674)  (291,0.00977836)  (292,0.00967125)  (293,0.00956534)  (294,0.00946068)  (295,0.00935718)  (296,0.00925491)  (297,0.00915377)  (298,0.00905384)  (299,0.008955)  (300,0.00885735)  (301,0.00876076)  (302,0.00866535)  (303,0.00857096)  (304,0.00847773)  (305,0.00838549)  (306,0.00829439)  (307,0.00820425)  (308,0.00811524)  (309,0.00802716)  (310,0.00794019)  (311,0.00785412)  (312,0.00776914)  (313,0.00768504)  (314,0.00760201)  (315,0.00751983)  (316,0.0074387)  (317,0.0073584)  (318,0.00727914)  (319,0.00720068)  (320,0.00712324)  (321,0.00704658)  (322,0.00697091)  (323,0.00689601)  (324,0.00682209)  (325,0.0067489)  (326,0.00667668)  (327,0.00660518)  (328,0.00653461)  (329,0.00646475)  (330,0.00639581)  (331,0.00632755)  (332,0.0062602)  (333,0.00619351)  (334,0.00612771)  (335,0.00606254)  (336,0.00599825)  (337,0.00593458)  (338,0.00587177)  (339,0.00580956)  (340,0.00574819)  (341,0.00568741)  (342,0.00562744)  (343,0.00556805)  (344,0.00550945)  (345,0.00545141)  (346,0.00539416)  (347,0.00533745)  (348,0.0052815)  (349,0.00522607)  (350,0.0051714)  (351,0.00511723)  (352,0.00506379)  (353,0.00501085)  (354,0.00495863)  (355,0.00490688)  (356,0.00485583)  (357,0.00480525)  (358,0.00475535)  (359,0.0047059)  (360,0.00465711)  (361,0.00460876)  (362,0.00456107)  (363,0.00451379)  (364,0.00446715)  (365,0.00442092)  (366,0.00437531)  (367,0.0043301)  (368,0.00428549)  (369,0.00424127)  (370,0.00419763)  (371,0.00415437)  (372,0.00411168)  (373,0.00406935)  (374,0.00402759)  (375,0.00398616)  (376,0.00394529)  (377,0.00390475)  (378,0.00386475)  (379,0.00382507)  (380,0.00378591)  (381,0.00374706)  (382,0.00370872)  (383,0.00367069)  (384,0.00363314)  (385,0.00359589)  (386,0.00355912)  (387,0.00352264)  (388,0.00348662)  (389,0.00345087)  (390,0.00341559)  (391,0.00338056)  (392,0.00334598)  (393,0.00331165)  (394,0.00327776)  (395,0.00324412)  (396,0.00321089)  (397,0.00317791)  (398,0.00314533)  (399,0.00311298)  (400,0.00308104)  (401,0.00304932)  (402,0.00301798)  (403,0.00298687)  (404,0.00295613)  (405,0.0029256)  (406,0.00289544)  (407,0.00286548)  (408,0.00283588)  (409,0.00280648)  (410,0.00277743)  (411,0.00274856)  (412,0.00272005)  (413,0.00269171)  (414,0.00266371)  (415,0.00263588)  (416,0.00260838)  (417,0.00258106)  (418,0.00255405)  (419,0.00252721)  (420,0.00250068)  (421,0.00247431)  (422,0.00244825)  (423,0.00242234)  (424,0.00239673)  (425,0.00237127)  (426,0.00234611)  (427,0.00232109)  (428,0.00229637)  (429,0.00227178)  (430,0.00224747)  (431,0.00222331)  (432,0.00219942)  (433,0.00217566)  (434,0.00215218)  (435,0.00212882)  (436,0.00210574)  (437,0.00208278)  (438,0.00206008)  (439,0.00203751)  (440,0.0020152)  (441,0.00199301)  (442,0.00197107)  (443,0.00194926)  (444,0.00192769)  (445,0.00190624)  (446,0.00188503)  (447,0.00186395)  (448,0.0018431)  (449,0.00182237)  (450,0.00180187)  (451,0.00178149)  (452,0.00176134)  (453,0.0017413)  (454,0.0017215)  (455,0.0017018)  (456,0.00168233)  (457,0.00166298)  (458,0.00164384)  (459,0.00162482)  (460,0.00160601)  (461,0.00158731)  (462,0.00156884)  (463,0.00155047)  (464,0.00153231)  (465,0.00151426)  (466,0.00149642)  (467,0.00147869)  (468,0.00146117)  (469,0.00144376)  (470,0.00142655)  (471,0.00140945)  (472,0.00139255)  (473,0.00137576)  (474,0.00135917)  (475,0.00134269)  (476,0.0013264)  (477,0.00131022)  (478,0.00129424)  (479,0.00127837)  (480,0.00126268)  (481,0.00124711)  (482,0.00123172)  (483,0.00121644)  (484,0.00120135)  (485,0.00118637)  (486,0.00117157)  (487,0.00115688)  (488,0.00114237)  (489,0.00112797)  (490,0.00111375)  (491,0.00109963)  (492,0.0010857)  (493,0.00107186)  (494,0.00105821)  (495,0.00104466)  (496,0.00103129)  (497,0.00101802)  (498,0.00100493)  (499,0.000991935)  (500,0.000979116)
};
\addlegendentry{$k=0$}

\addplot [\revcolor,densely dotted]
coordinates {
(0,1)
(4.31246405814,0.0139658466039531)
(8.62492811628,0.0041668541398281)
(12.93739217442,0.000847698156070637)
(17.24985623256,0.000158136379991418)
(21.5623202907,3.47168021764028e-05)
(25.87478434884,1.58657865524238e-05)
(30.187248406979997,6.49958511965696e-06)
(34.49971246512,4.22015623115358e-06)
(38.81217652326,2.57511691325517e-06)
(43.1246405814,1.40125091697247e-06)
(47.437104639539996,6.16093768733579e-07)
(51.74956869768,3.71781289191082e-07)
(56.06203275582,1.53726633613046e-07)
(60.37449681395999,6.52565438202897e-08)
(64.6869608721,3.28778752255404e-08)
(68.99942493024,1.95846870722275e-08)
(73.31188898837999,1.79699504778845e-08)
(77.62435304652,1.7403260955763e-08)
(81.93681710466,1.43558186307187e-08)
(86.2492811628,7.78242754790345e-09)
(90.56174522094,5.32391815822228e-09)
(94.87420927907999,3.83731158071241e-09)
(99.18667333722,1.62082917304287e-09)
(103.49913739536,5.07374790297855e-10)
(107.81160145349999,1.98648892242197e-10)
(112.12406551164,7.32833323817348e-11)
(116.43652956977999,2.08190482596783e-11)
(120.74899362791999,6.22919571436964e-12)
(125.06145768606,1.76420320026259e-12)
(129.3739217442,5.82468868624697e-13)
(133.68638580234,2.04945468106169e-13)
(137.99884986048,4.66347260525612e-14)
};
\addlegendentry{$k=1$}

\addplot [\revcolor,loosely dashed]
coordinates {
(0,1)
(7.62492811628,0.0105245937251093)
(15.24985623256,0.00219980728657953)
(22.87478434884,0.00033474466731038)
(30.49971246512,3.41878035950007e-05)
(38.1246405814,5.00048011464107e-06)
(45.74956869768,2.29402673953959e-06)
(53.37449681395999,8.17760436119531e-07)
(60.99942493024,5.40246140746962e-07)
(68.62435304652,3.60361513215292e-07)
(76.2492811628,1.40351193592656e-07)
(83.87420927907999,7.37327241828337e-08)
(91.49913739536,3.76840067690339e-08)
(99.12406551164,1.62467035686949e-08)
(106.74899362791999,7.20099839149835e-09)
(114.37392174419999,2.46373963882223e-09)
(121.99884986048,1.07636865373654e-09)
(129.62377797675998,9.82222933846704e-10)
(137.24870609304,9.54419770641204e-10)
(144.87363420932,7.85662451050229e-10)
(152.4985623256,2.81235492222685e-10)
(160.12349044188,9.75578758138094e-11)
(167.74841855815998,6.58084475902612e-11)
(175.37334667444,4.2905015390462e-11)
(182.99827479072,2.00802355450872e-11)
(190.62320290699998,8.14921396691505e-12)
(198.24813102328,2.88791740160256e-12)
(205.87305913955998,5.07567626264034e-13)
(213.49798725583997,6.68390795775875e-14)
};
\addlegendentry{$k=2$}

\addplot [\revcolor,dashed]
coordinates {
(0,1)
(17.5623202907,0.0102042983778133)
(35.1246405814,0.00199633712700245)
(52.6869608721,0.000247919901412568)
(70.2492811628,1.91691100717472e-05)
(87.8116014535,1.00340811333683e-06)
(105.3739217442,4.2766858142064e-08)
(122.9362420349,9.60396897586651e-09)
(140.4985623256,4.63042467611712e-09)
(158.0608826163,2.50863808804697e-09)
(175.623202907,1.77144525432458e-09)
(193.1855231977,7.94439552478128e-10)
(210.7478434884,2.11831330952732e-10)
(228.3101637791,9.7102397306046e-11)
(245.8724840698,1.7146375391556e-11)
(263.4348043605,7.68907478564174e-12)
(280.9971246512,4.64025568958641e-12)
(298.5594449419,1.5596522324084e-12)
(316.1217652326,6.68147643877975e-13)
(333.6840855233,5.41368350608748e-13)
(351.246405814,5.25496491148563e-13)
(368.8087261047,4.7110174511416e-13)
(386.3710463954,2.5693445967383e-13)
(403.9333666861,1.00997280049704e-13)
(421.4956869768,3.24515601161827e-14)
};
\addlegendentry{$k=5$}

\addplot [\revcolor]
coordinates {
(0,1)
(34.1246405814,0.0102031443864273)
(68.2492811628,0.00199595424384562)
(102.37392174419999,0.000247782434939009)
(136.4985623256,1.91485716863828e-05)
(170.623202907,1.00149445993677e-06)
(204.74784348839998,4.04356140970216e-08)
(238.87248406979998,2.73590064131848e-09)
(272.9971246512,2.12281803016104e-10)
(307.1217652326,5.86538701950592e-12)
(341.246405814,6.95539829793314e-13)
(375.37104639539996,4.74299006171461e-13)
(409.49568697679996,2.61968391539119e-13)
(443.62032755819996,1.51058975156e-13)
(477.74496813959996,3.69633947899524e-14)
};
\addlegendentry{$k=10$}

\end{semilogyaxis}

\end{tikzpicture}}
\caption{Typical residual behavior for MINRES, applied to the
problem $J^{(h)}(\psi^{(h)}) \phi^{(h)} = b^{(h)}$, preconditioned with $R_k^{(h)}(\psi^{(h)})$
for different $k$, with each $\psi^{(h)}\equiv b^{(h)}\equiv 1$,
initial guess $\phi^{(h)}_0\equiv 0$, for $\Omega^{(h)}_{\text{sq}}$ in $\A_z$.
The number of unknowns is $1000^2$ in all cases.
10 AMG steps solve the preconditioning problem in each MINRES step up to at least $\|r_{\text{p}}\|\le 10^{-12}$.
\rev{In the figure on the right,
the computational complexity is measured in terms of the cost of one matrix-vector
multiplication. For this setting, the application of one V-cycle costs as much as about $3.31$ matrix-vector
multiplications.}
\label{fig:v-cycle-preconditioning}}
\end{figure}

While $R^{(h)}_{\infty}(\psi^{(h)})$ is obviously more expensive to apply,
it is expected that it will yield a smaller number of Krylov iterations as compared
to preconditioning with $R^{(h)}_{1}(\psi^{(h)})$. Figure~\ref{fig:v-cycle-preconditioning}
illustrates this: For a fixed setup, the preconditioners
$R_k^{(h)}(\psi^{(h)})$ with $k\in\{1,2,5,10\}$ are compared,
where for this particular case $R_{10}^{(h)}(\psi^{(h)})\approx R^{(h)}_{\infty}(\psi^{(h)})$ to machine-precision.
Preconditioning with $R_{10}^{(h)}(\psi^{(h)})$ indeed results in the smallest
number of required MINRES iterations;
if fewer V-cycles are applied per iteration, the number of iterations increases.
\rev{
A better measure for the overall computational cost
than the sheer number of Krylov iterations, however, is
the number of performed V-cycles together with the matrix-vector products.
While the latter mainly depends on the number of nonzeros in the kinetic energy operator $K^{(h)}$,
the cost of the former also depends the many parameters of AMG. In all of the experiments
performed in this paper, the cost of the application of one V-cycle is between three and four
times the cost of a matrix-vector product of the corresponding matrix.}
As can be seen in the right panel of
figure~\ref{fig:v-cycle-preconditioning}, \rev{no more than
the equivalent of about 140 matrix-vector products is} are
required in total to converge the MINRES process in
combination a single V-cycle preconditioning.
At the same time, 10 cycles per step require
\rev{the equivalent of about 480 matrix-vector multiplications}.
This points to the fact that the approximate
inversion with a single V-cycle will lead to the fastest solver.

For the experiments in figure~\ref{fig:v-cycle-preconditioning} and
all experiments in the remainder of this paper, smoothed-aggregation AMG
with one pre- and one post-smoothing step of \rev{symmetric} Gauss--Seidel was used.
The method is implemented using PyAMG~\cite{BeOlSc2008}.

We now look at the application to the two-dimensional regular polygons in the $x$-$y$-plane
\begin{align*}
&\Omega_{\text{tri}}\dfn H_{\text{convex}}\left(\left\{\begin{pmatrix}0\\5\end{pmatrix}, \begin{pmatrix}-5\sqrt{3}/2\\-5/2\end{pmatrix}, \begin{pmatrix}5\sqrt{3}/2\\-5/2\end{pmatrix}\right\}\right) \text{ (figure~\ref{subfig:tri})},\\
&\Omega_{\text{sq}}\dfn \{\x: \left\|\x\right\|_{\infty}<5/\sqrt{2}\} \text{ (figure~\ref{subfig:rect})},\\
&\Omega_{\text{circ}}\dfn \{\x: \left\|\x\right\|_2<5\} \text{ (figure~\ref{subfig:circ})},
\end{align*}
and the three-dimensional regular polyhedra
\begin{align*}
&\Omega_{\text{tet}}\dfn 5 \cdot H_{\text{convex}}\left(\left\{\begin{pmatrix}0\\0\\1\end{pmatrix}, \begin{pmatrix}2\sqrt{2}/3\\0\\-1/3\end{pmatrix}, \begin{pmatrix}-\sqrt{2}/3\\\sqrt{2/3}\\-1/3\end{pmatrix}, \begin{pmatrix}-\sqrt{2}/3\\-\sqrt{2/3}\\-1/3\end{pmatrix}\right\}\right) \text{ (figure~\ref{subfig:tet})},\\
&\Omega_{\text{cube}}\dfn \{\x: \left\|\x\right\|_{\infty}<5/\sqrt{3}\} \text{ (figure~\ref{subfig:cube})},\\
&\Omega_{\text{ball}}\dfn \{\x: \left\|\x\right\|_2<5\} \text{ (figure~\ref{subfig:ball})},
\end{align*}
all centered at the origin with circumradius 5.
For each domain, both the potentials
\begin{equation}\label{eq:mvp0}
\A_z(\x)\dfn \tfrac{1}{2}(-y, x, 0)^\tp,
\end{equation}
representing the homogeneous field $\B=(0,0,1)^\tp$, and
\[
\A_{\text{d}}(\x)\dfn \tfrac{1}{\|\x-\x_0\|^{3}} (\bm{m}\times(\x-\x_0)),
\]
representing the inhomogeneous field generated by a magnetic dipole at
the location
\[
\x_0 =
\begin{cases}
(0,0,1)^\tp\text{ for the 2D domains,}\\
(0,0,6)^\tp\text{ for the 3D domains,}\\
\end{cases}
\]
and with the dipole moment $\bm{m}=(0,0,1)^\tp$, are considered.
For all experiments, we considered $J^{(h)}(\psi^{(h)}))$ with
$\psi^{(h)}\equiv 1$.  For other choices of $\psi^{(h)}$
see the paragraph on numerical continuation below.

\begin{figure}
\centering
\setlength\figurewidth{0.5\textwidth}
\setlength\figureheight{0.4\textwidth}
\subfloat[Tetrahedron in $\A_z$.]{%
\begin{tikzpicture}

\begin{semilogxaxis}[
xlabel={dimension of solution space},
ylabel={number of iterations},
xmin=250, xmax=130000,
ymin=1, ymax=40,
width=\figurewidth,
height=\figureheight
]
\addplot [dotted]
coordinates {
(406,75)
};

\addplot [densely dotted, mark=o, mark options={solid}]
coordinates {
(406,20)
(3018,22)
(9676,22)
(22270,22)
(42748,22)
(71832,22)
(112580,22)
};

\addplot [mark=triangle]
coordinates {
(406,20)
(3018,19)
(9676,19)
(22270,18)
(42748,18)
(71832,18)
(112580,18)
};

\end{semilogxaxis}

\end{tikzpicture}}
\subfloat[Tetrahedron in $\A_{\text{d}}$.]{%
\begin{tikzpicture}

\begin{semilogxaxis}[
xlabel={dimension of solution space},
yticklabels=\empty,
xmin=250, xmax=130000,
ymin=1, ymax=40,
width=\figurewidth,
height=\figureheight,
]
\addplot [dotted]
coordinates {
(406,91)
};

\addplot [densely dotted, mark=o, mark options={solid}]
coordinates {
(406,26)
(3018,29)
(9676,27)
(22270,29)
(42748,27)
(71832,28)
(112580,28)
};

\addplot [mark=triangle]
coordinates {
(406,26)
(3018,25)
(9676,23)
(22270,23)
(42748,22)
(71832,22)
(112580,22)
};

\end{semilogxaxis}

\end{tikzpicture}}\\
\subfloat[Cube in $\A_z$.]{%
\begin{tikzpicture}

\begin{semilogxaxis}[
xlabel={dimension of solution space},
ylabel={number of iterations},
xmin=250, xmax=130000,
ymin=1, ymax=40,
width=\figurewidth,
height=\figureheight
]
\addplot [dotted]
coordinates {
(250,14)
(2000,27)
(6750,40)
(16000,53)
(31250,66)
(54000,80)
(85750,93)
};

\addplot [densely dotted, mark=o, mark options={solid}]
coordinates {
(250,3)
(2000,18)
(6750,18)
(16000,19)
(31250,19)
(54000,18)
(85750,19)
(128000,18)
};

\addplot [mark=triangle]
coordinates {
(250,3)
(2000,10)
(6750,10)
(16000,10)
(31250,10)
(54000,10)
(85750,10)
(128000,10)
};

\end{semilogxaxis}

\end{tikzpicture}}
\subfloat[Cube in $\A_{\text{d}}$.]{%
\begin{tikzpicture}

\begin{semilogxaxis}[
xlabel={dimension of solution space},
yticklabels=\empty,
xmin=250, xmax=130000,
ymin=1, ymax=40,
width=\figurewidth,
height=\figureheight,
]
\addplot [dotted]
coordinates {
(250,15)
(2000,34)
(6750,44)
(16000,59)
(31250,74)
(54000,89)
};

\addplot [densely dotted, mark=o, mark options={solid}]
coordinates {
(250,6)
(2000,22)
(6750,29)
(16000,31)
(31250,31)
(54000,32)
(85750,32)
(128000,31)
};

\addplot [mark=triangle]
coordinates {
(250,6)
(2000,11)
(6750,10)
(16000,10)
(31250,10)
(54000,10)
(85750,10)
(128000,10)
};

\end{semilogxaxis}

\end{tikzpicture}}\\
\subfloat[Ball in $\A_z$.]{%
\begin{tikzpicture}

\begin{semilogxaxis}[
xlabel={dimension of solution space},
ylabel={number of iterations},
xmin=250, xmax=130000,
ymin=1, ymax=40,
width=\figurewidth,
height=\figureheight
]
\addplot [dotted]
coordinates {
(250,14)
(1222,70)
};

\addplot [densely dotted, mark=o, mark options={solid}]
coordinates {
(250,3)
(1222,20)
(8024,19)
(24248,19)
(54920,19)
(104160,18)
(176708,18)
(273598,18)
};

\addplot [mark=triangle]
coordinates {
(250,3)
(1222,18)
(8024,17)
(24248,16)
(54920,15)
(104160,14)
(176708,14)
(273598,14)
};

\end{semilogxaxis}

\end{tikzpicture}}
\subfloat[Ball in $\A_{\text{d}}$.]{%
\begin{tikzpicture}

\begin{semilogxaxis}[
xlabel={dimension of solution space},
yticklabels=\empty,
xmin=250, xmax=130000,
ymin=1, ymax=40,
width=\figurewidth,
height=\figureheight,
]
\addplot [dotted]
coordinates {
(250,15)
(1222,122)
};

\addplot [densely dotted, mark=o, mark options={solid}]
coordinates {
(250,6)
(1222,37)
(8024,34)
(24248,34)
(54920,32)
(104160,34)
(176708,32)
(273598,32)
};

\addplot [mark=triangle]
coordinates {
(250,6)
(1222,33)
(8024,29)
(24248,27)
(54920,27)
(104160,27)
(176708,27)
(273598,26)
};

\end{semilogxaxis}

\end{tikzpicture}}
\caption{MINRES performance for different three-dimensional domains and magnetic vector potentials
at the state $\psi_0\equiv 1$.
The plots show the number of iterations necessary to reach the relative residual
of $10^{-11}$ in the norm given by~(\ref{eq:disc inner2}) as a function of the dimension
of the problem. Starting guess for the linear iterations is $\phi_0\equiv 0$
throughout.
Plotted are results
for the unpreconditioned problem (\ref{addplot:unprec}),
the preconditioner $R_1^{(h)}(\psi_0^{(h)})$ (\ref{addplot:prec1}), and
the preconditioner $R_{\infty}^{(h)}(\psi_0^{(h)})$ (\ref{addplot:prec2}).}
\label{fig:preconditioning 3d}
\end{figure}
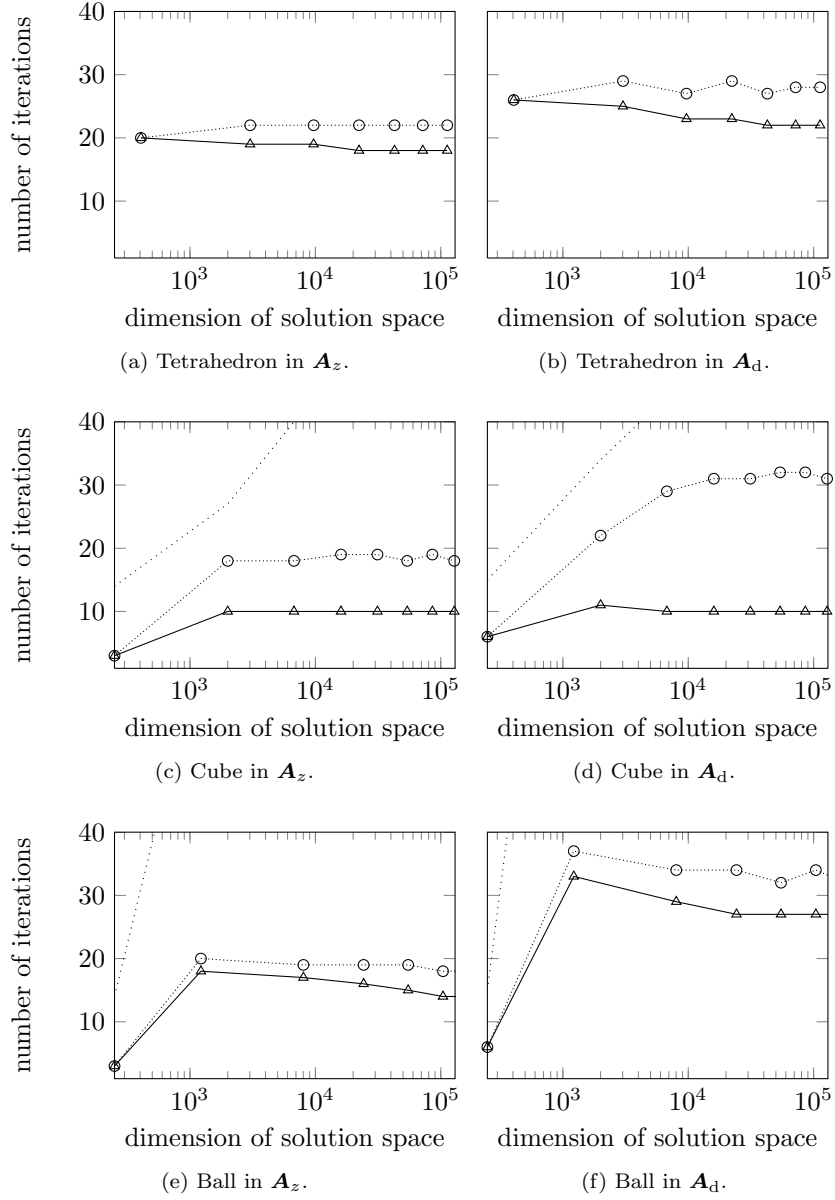

Figures~\ref{fig:preconditioning 2d} and~\ref{fig:preconditioning 3d} show the number of MINRES
iterations as a function of the dimension of the solution space.
For the unpreconditioned system, the
number of iteration increases as expected since the finer discretization makes the
condition number of $K^{(h)}$ and hence $J^{(h)}$ larger.
In contrast to this, when $R^{(h)}_1$ and $R^{(h)}_{\infty}$ are applied as
preconditioners, the number of iterations remains bounded in all problem
settings as the discretization refines.
Although the
number of iterations, when preconditioned with $R^{(h)}_1$, is slightly larger
compared to preconditioning
with $R_{\infty}^{(h)}$, the former is actually computationally cheaper as discussed above
(see figure~\ref{fig:v-cycle-preconditioning}).
These numerical experiments suggest that
for various fixed domains and magnetic vector potentials,
the number of iterations of the Krylov solver is independent of the number of
unknowns.

\begin{figure}
\centering
\setlength\figurewidth{0.5\textwidth}
\setlength\figureheight{0.4\textwidth}
\subfloat[Triangle in $\A_z$.]{%
%
}
\caption{MINRES performance for different two-dimensional domains and magnetic vector potentials
at the state $\psi_0\equiv 1$.
The plots show the number of iterations necessary to reach the relative residual
of $10^{-11}$ in the norm given by~(\ref{eq:disc inner2}) as a function of the dimension
of the problem. Starting guess for the linear iterations is $\phi_0\equiv 0$
throughout.
Plotted are results
for the unpreconditioned problem (\ref{addplot:unprec}),
the preconditioner $R_1^{(h)}(\psi_0^{(h)})$ (\ref{addplot:prec1}), and
the preconditioner $R_{\infty}^{(h)}(\psi_0^{(h)})$ (\ref{addplot:prec2}).}
\label{fig:preconditioning 2d}
\end{figure}

\paragraph{Numerical parameter continuation}
A common application of the Newton--Krylov solver is in a numerical
continuation context where a family (or curve) of solution states (from a given state space $X$)
is constructed as a function of a parameter (from a given parameter space $A$)
in the system. This is a popular way of exploring
the solution landscape of nonlinear equations,
and amongst the most widely used algorithms for this purpose is
pseudo-arclength continuation \cite{krauskopf2007}, a
predictor-corrector method. Here, in each continuation step
an initial guess is constructed
as an extrapolation to the solution curve in $X\times A$
which is then corrected perpendicularly to the extrapolation,
typically involving a Newton--Krylov process.
In applications, many curves, each with thousands of
solutions, are computed. Each continuation step requires the solution of
a nonlinear system, each of which requires the solution of a Jacobian~system.

\begin{figure}
\centering
\setlength\figurewidth{0.45\textwidth}
\setlength\figureheight{0.8\figurewidth}
\subfloat[][Gibbs energy as a function of $\mu$. Note the so-called \emph{swallow tails} which
are typical for numerical parameter continuation. Structural transitions
in the solutions usually happen around here.]{%
}
\hspace{2ex}
\subfloat[][Number of \rev{MINRES} iterations till $\|r\|\le 10^{\rev{-8}}$ in the last Newton step as a function of $\mu$.
\rev{With increasing energy, stability is lost
and the presence of unstable eigenmodes
hampers the performance of the preconditioner.}%
]{%
%
\label{subfig:sizemu10iters}}\\
\setlength\figurewidth{0.3\textwidth}
\subfloat[][Solution $|\psi|^2$, $\arg\psi$ at continuation step 22 with
$\mu\approx 0.47$, $F^{(h)}\approx-0.47$. The system has undergone a
first transition and four vortices have moved in.]{%
\includegraphics{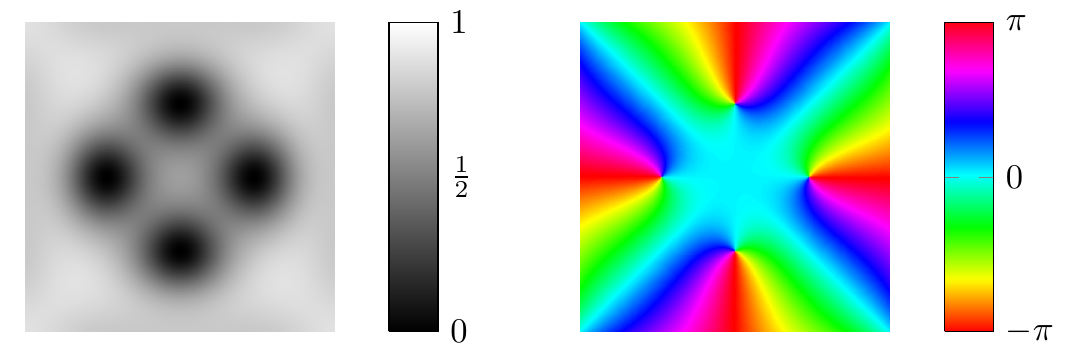}
\label{fig:step22}}\\
\subfloat[][Solution $|\psi|^2$, $\arg\psi$ at continuation step 49 with
$\mu\approx 0.93$, $F^{(h)}\approx-0.12$. The system has undergone a second
transition and now contains eight vortices.]{%
\includegraphics{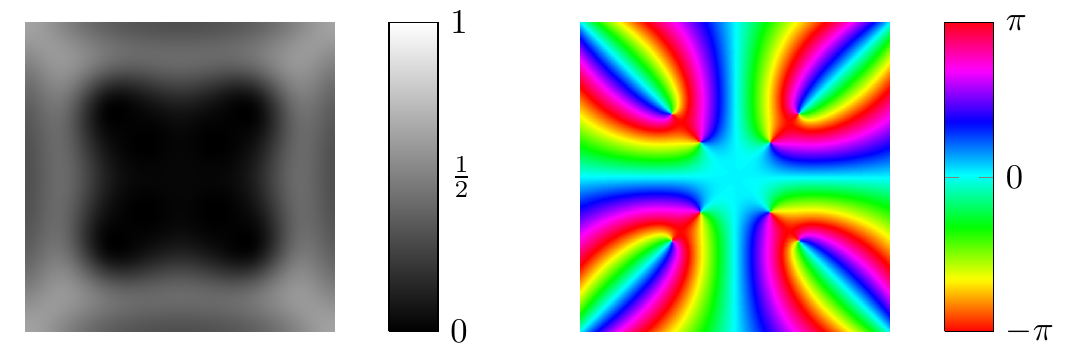}
\label{fig:step49}}
\caption{The performance of the preconditioned Krylov solver in the
context of numerical continuation in $\mu \A_z$ for $\Omega^{(h)}_{\text{sq}}$
with edge length $10$, $1000^2$ unknowns.
The solution is continued in the parameter $\mu$ with
the help of pseudo arc-length continuation where each continuation step
requires the solution of a nonlinear system.
}
\label{fig:continuation_size}
\end{figure}

As this setting presents a typical use case for the preconditioner introduced
above, a representative problem is discussed in this section.
As opposed to all previous numerical experiments, the state $\psi^{(h)}$
deviates significantly from the initial state $\psi^{(h)}_0\equiv 1$
in the corresponding numerical experiment (see figures~\ref{fig:step22}, \ref{fig:step49}).

Figure~\ref{fig:continuation_size}, illustrates the performance of
the $R^{(h)}_{\rev{1}}$-preconditioned Krylov--solver for
$\Omega^{(h)}_{\text{sq}}$ with edge length $10$ in $\mu \A_z$ (\ref{eq:mvp0}), $\mu\in\R$.
The strength $\mu$ of the magnetic field is used as continuation parameter,
and the continuation is started with the trivial solution $\psi^{(h)}_0\equiv 1$
at $\mu = 0$. As $\mu$ increases, the solution starts to deviate
from the homogeneous state.
Throughout the parameter continuation, vortices appear in the domain and the state loses
its stability \cite{SAV:2012:NBS}, a process marked by eigenvalues
of the Jacobian crossing the origin, i.e., a change of definiteness of the
Jacobian operator.
The right panel of figure~\ref{fig:continuation_size}
shows, for each point on the continuation curve, the number of
iterations in the preconditioned \rev{MINRES} solver that was required to solve the
Jacobian system up to $\|r\|\le 10^{-8}$ in the last Newton step.
\rev{While initially around 50 iterations are required, the introduction
of an unstable eigenvalue at the swallow tail ($\mu \approx 0.30$)
slows down MINRES convergence. This is due to the fact that the
positive-definite preconditioner does not capture negative eigenvalues.
While the convergence is not slowed down by an order of magnitude,
the highly unstable high-energy states around $\mu= 0.8$ require
up to 150 MINRES iterations to converge.
The local peaks in the iteration requirements are due to
the inherent loss of orthogonality of Krylov basis vectors
in MINRES \cite{SleVM00}. This effect can be alleviated
by storing and fully reorthogonalizing the Krylov basis
in each MINRES step such as implemented in GMRES.}

\section{Conclusions}
The authors propose a preconditioned Newton--Krylov algorithm that
solves the extreme type-II Ginzburg--Landau equation. The solution
method uses an AMG preconditioning strategy that yields optimal
convergence and scalability for mesoscopic domains.

The Ginzburg--Landau operator consists of a kinetic energy
operator that depends on the given magnetic field and a nonlinear
reaction term.
The linearization of the operator is analyzed and it is
found that the Jacobian of the system is self-adjoint with respect
to the nonstandard inner product~\eqref{eq:scalar product}.
Its spectrum
is indefinite if $\psi$ describes a physically unstable solution of the
equation.
The properties of the kinetic energy operator $K$ are also
discussed and it is found to be self-adjoint and positive
(semi)-definite. These properties are
maintained after discretization with finite volumes and link
variables.
The proposed preconditioner takes advantage of this by applying
an algebraic multigrid scheme to the operator $P^{(h)}(\psi^{(h)}) = K^{(h)} + 2|\psi^{(h)}|^2$.
Numerical results for
representative domains point towards the optimality of the algorithm
in the sense of independence of the number of linear solver iterations
from the discretization resolution. This suggests that, qualitatively, no
further improvement can be reached.

Moreover, the performance of the preconditioner is assessed in a
numerical parameter continuation context where
a family of solutions is generated for
changing strength of the applied magnetic field.
The good convergence results from the test domains are confirmed here.
The presence of negative eigenvalues, however, slows down the Krylov
convergence if used with a CG solver (figure~\ref{subfig:sizemu10iters}).
Moreover, other factors, such as a large domain size, have shown to hamper the
convergence. To gain deeper insight into the convergence
behaviors, clearer results than (\ref{eq:lambda})
on the spectrum of $(P^{(h)})^{-1}J^{(h)}$ are needed.

Nevertheless, this research opens up new possibilities for the exploration of the
energy landscape of type-II superconductors. Computation of
three-dimensional problems are now accessible with grid resolutions on
par with current two-dimensional calculations.

A natural extension of the presented work is to approach the solution
of the full Ginzburg--Landau problem in which the magnetic vector
potential cannot be treated as given \cite{DGP:1992:AAG}. Numerous
numerical and computational challenges are posed there, e.g.,
how to efficiently solve the Jacobian system.
The presented preconditioner could be used to construct a block-preconditioning
strategy for the general (nonextreme-type-II) Ginzburg--Landau equations.

\section*{Acknowledgments}
We acknowledge fruitful discussions with Qiang Du, Andrew G. Sa\-lin\-ger, Gregory D. Sjaardema, and Mark Hoemmen.
We are also grateful to the Research Foundation Flanders (FWO) for financial support through the project G017408N.

\bibliographystyle{elsarticle-num}
\bibliography{gl}

\end{document}